\documentclass[english,twocolumn,aps,prd,longbibliography,crop,tikz,superscriptaddress]{revtex4-2}
\usepackage[english]{babel}
\usepackage[utf8]{inputenc}
\usepackage{amsfonts}
\usepackage{amsmath}
\usepackage{afterpage}
\usepackage{graphicx}
\usepackage{listings}
\usepackage{xcolor}  
\usepackage{bbold}
\graphicspath{ {./images/} }
\usepackage[colorinlistoftodos, color=green!40, prependcaption]{todonotes}
\usepackage{amsthm}
\usepackage{mathtools}
\usepackage{physics}
\usepackage{xcolor}
\usepackage{graphicx}
\usepackage[left=23mm,right=13mm,top=35mm,columnsep=15pt]{geometry} 
\usepackage{adjustbox}
\usepackage{placeins}
\usepackage[T1]{fontenc}
\usepackage{lipsum}
\usepackage{csquotes}
\usepackage[pdftex, pdftitle={Article}, pdfauthor={Author}]{hyperref} 
\begin{document}
\title{QGeo: A Python package for calculating geodesic control functions for quantum computing}

\author{Sean T. Crowe}
\thanks{Corresponding Author: sean.t.crowe2.civ@us.navy.mil}
\affiliation{Naval Information Warfare Center Pacific, San Diego, CA, 92152, United States}
\author{Joshua J. Leiter}
\affiliation{Naval Information Warfare Center Pacific, San Diego, CA, 92152, United States}
\author{John P. T. Stenger}
\affiliation{U.S. Naval Research Laboratory, Washington, DC 20375, United States}
\author{ Zachary L. Barvian}
\affiliation{Naval Information Warfare Center Pacific, San Diego, CA, 92152, United States}
\author{Joseph A. Diaz}
\affiliation{Naval Information Warfare Center Pacific, San Diego, CA, 92152, United States}
\author{Shoshana Krishel}
\affiliation{San Diego State University, San Diego, CA, 92152, United States}
\author{Joanna N. Ptasinski}
\affiliation{Naval Information Warfare Center Pacific, San Diego, CA, 92152, United States}
\author{Daniel Gunlycke}
\affiliation{U.S. Naval Research Laboratory, Washington, DC 20375, United States}

\begin{abstract}
We present a new Python package that uses the established notion of geometric quantum complexity to numerically compute the difficulty associated with preparing a given unitary transformation on a quantum computer. The numerical procedure we implement is presented and discussed. Analyzed quantum circuits include: the quantum fourier transform for up to four qubits, a random circuit with depth 100, and a circuit for analyzing the evolution of a fermionic chain with several lattice sites. This package can be found for download at \url{https://github.com/JAGDiaz/quantum-geodesics}.
\end{abstract}

\keywords{quantum information, geometry, complexity}

\maketitle

\section{Introduction} \label{sec:intro}

    Quantum computers implement unitary transformations on quantum states. In practice, this is done by decomposing a target unitary transformation into a product of elementary unitary transformations called quantum gates such that the target unitary transformation can be expressed as
    
    \begin{equation}
        \label{eq:Target Unitary}
        U_{\rm T}=g_N...g_1.
    \end{equation}
    Here the $g_{i}$'s are elementary quantum gates. The specific set of elementary gates which are available will depend on the specific quantum computer being used.
    At the hardware level, the implementation of these gates is specialized\footnote{For example implementations of elementary gates on superconducting qubits see \cite{Krantz_2019}}, and most mainstream platforms for quantum computation usually have a relatively small set of elementary gates available. For example, the IBM quantum computer Torino can only implement tensor products of: the identity matrix, an X gate, an SX gate, an Rz gate, and a controlled-Z gate \cite{ibmquantum}. 
    
    The number of gates required to implement a target unitary is called the gate complexity \cite{Brown_2019}. Given an elementary gate set, a quantum processor, and a target unitary transformation, the problem of finding an implementation of the target unitary that minimizes the complexity is called the quantum circuit mapping problem, which is known to be difficult \cite{Wille}.

    An alternative measure of complexity of a unitary operator is the geodesic quantum complexity \cite{geo_quant}. In this approach, the difficulty is associated with preparing a given unitary matrix specified by a metric on the unitary group and then looking for curves of minimal length that connect the identity operator to the target unitary matrix. These minimal length curves are called geodesics and the length of the minimal geodesic is then a measure of complexity.  

    Concretely, the evolution of the quantum state is generated by a Hamiltonian $\hat{H}(t)$. This implies a unitary evolution that is given by
    
    \begin{equation}\label{general_solution}
        \hat{U}(t)=\mathcal{T} \mathrm{e}^{-i \int_{0}^{t}dt' \hat{H}(t')},
    \end{equation}
     $\mathcal{T}$ is the time-ordering operator, which orders operators in descending time order such that operators evaluated at the latest times are on the left. In \eqref{solution} it is shown that this is a solution to the Schr\"odinger equation.

    Onto \eqref{general_solution}  we impose the boundary conditions: $\hat U(0)=\mathbb{1}$ and $\hat U(T)=U_{\mathrm{T}}$. This evolution defines a path, $\gamma$, in the unitary group, and the length of that path is specified by a phenomenologically motivated real-valued, symmetric, inner product defined by
    
    \begin{equation}
        \label{metric}
        \langle \hat{H},\hat{J} \rangle=\frac{1}{2^{n}}\mathrm{tr}\left(\hat{H}\mathcal{G}\left(\hat{J}\right)\right),
    \end{equation}
    where $n$ is the number of qubits and the map $\mathcal{G}$ is a super-operator defined in equation \eqref{eq:FG} that penalizes operations that have weight greater than two.  This is meant to mimic actual lab conditions where the implementation of three or more qubit gates is difficult. The length of the path is given by \eqref{complexity}. The length of the path that minimizes \eqref{complexity} is interpreted as the complexity of the target unitary.
    
    \begin{equation}\label{complexity}
        C[\gamma]=\int_{0}^{T}\langle \hat{H}(t),\hat{H}(t)\rangle^{1/2}dt
    \end{equation}

    Paths on which this complexity is stationary are called geodesics. While geodesic paths may not be global minima of the complexity, the global minimum will correspond to a geodesic path. Since its inception, the notion of geodesic complexity has been applied in broad spectrum of areas which span quantum information to quantum gravity \cite{Brown_2019,chowdhury2024upperboundsquantumcomplexity,Hackl_2018,Camargo_2021,Chandra_2021,Flory_2019,Flory_2020,Flory_2020-2}.
     
     Part of the original motivation for formulating the notion of geodesic quantum complexity was to put lower bounds on the exact gate complexity, and an upper bound on the approximate gate complexity associated with preparing $U$. To our knowledge, however, it has not been applied to its original purpose of bounding gate complexities of operationally relevant unitary matrices. The purpose of this article it to apply the notion of geometric complexity to several examples, including the quantum Fourier transform. In Sec. \ref{sec:Approach}, the problem is defined more concretely and strategy for solving it is discussed. In Sec. \ref{sec:AA}, analytic approximations to the problem are discussed. The numerical implementation of the strategy and specific calculations of geodesic complexities are then presented in Sec. \ref{sec:NA}.
     
\section{Approach} \label{sec:Approach}

    In this section, we discuss a numerical approach for calculating the quantum complexity of a given unitary operator. Essentially, our problem consists of calculating solutions of the Schr\"odinger-geodesic equation, which is a coupled set of ordinary differential equations which are discussed below. Our solutions are subject to the boundary conditions $U(0)=\mathbb{1}$, and $U(T)=U_{\mathrm{T}}$. This is a boundary value problem. A standard tool for numerically solving this kind of problem is shooting \cite{shooting}. However, because the dimensionality of our problem is very high, shooting is no longer effective and other approaches must be used. We approach this problem as in \cite{geo_quant}. To keep this report self-contained we reproduce details of this method here.
    
 The Schr{\"o}dinger-geodesic equation is derived by Dowling and Nielsen \cite{geo_quant} and is given by:

    \begin{align}
    \begin{split}
        \label{eq:Schrodinger-geodesic}
        \dot{U}\left(t\right)+\frac{i}{\hbar}H\left(t\right)U\left(t\right)&=0\\
        \dot{H}+i\mathcal{F}\left(\left[H,\mathcal{G}\left(H\right)\right]\right)&=0,\\
        \end{split}
    \end{align}
    Again, subject to the boundary conditions: $U(0)=\mathbb{1}$, and $U(T)=U_{\mathrm{T}}$. The solutions of this system of equations will be curves on $SU\left(2^{n}\right)$. Moreover, solving this system will give us the tangent vectors/Hamiltonian $H$ that can then be used in \eqref{complexity} to calculate the complexity of this path.  The super-operator $\mathcal{G}$, and its inverse super-operator $\mathcal{F}$ are given in equation \ref{eq:FG}.

    \begin{align}
        \label{eq:FG}
        \mathcal{G}=\mathcal{P}+q\mathcal{Q}\\
        \nonumber \mathcal{F}=\mathcal{P}+q^{-1}\mathcal{Q}
    \end{align}

     $\mathcal{P}$ is the identity map on generators of one and two qubit operations and the zero map on generators of three and higher qubit operations. The map $\mathcal{Q}$ is the zero map on one and two qubits, and the identity map on three and higher qubits. The parameter $q$ is a factor which penalizes operations corresponding to more than two qubits. This is to mimic lab conditions. $q=4^{n}$ is a regime of computational interest \cite{geo_quant}.
    
    We set the initial condition in equation \ref{eq:Schrodinger-geodesic} to be $H_{0}=V_{0}^{i}\sigma_{i}$, where $V_{0}^{i}$ are real constants, and $i\sigma_{i}$ is a basis for $\mathfrak{su}\left(2^{n}\right)$. The $\sigma_{i}$ are constructed from taking $n-1$ Kronecker products of the Pauli matrices in $\mathfrak{su}\left(2\right)$. 

    \subsection{Solution Strategy}

    Having stated the problem, we discuss the strategy for constructing a solution to the boundary value problem specified at \eqref{eq:Schrodinger-geodesic}. Given a solution to the problem for a given penalty factor, $q$, it is possible to construct the solution for, $q+\Delta$ when $\Delta$ is small enough. This is because for this regime, the problem can be linearized, and then analyzed more easily. Using this perturbative approach, a solution to \eqref{eq:Schrodinger-geodesic} can be constructed for arbitrarily large penalty factor, including the regime where $q\approx 4^n$, corresponding to more realistic lab conditions. The base case for this procedure is the $q=1$ case. As will be discussed later, this case can be solved analytically, and therefore serves as a good starting point. Given a solution to \eqref{eq:Schrodinger-geodesic} for a given q, $U_q(t)$, we model the solution at $q+\Delta$ as:
    \begin{equation}
        U_{q+\Delta}=U_q(t)\mathrm{e}^{-i \Delta J(t)}
    \end{equation}
   This implies a perturbation to the Hamiltonian which can be found by computing the quantity: $i \dot{U}U^{\dagger}$. This is found to be,
   \begin{equation}\label{perturbed_ham}
   \begin{split}
       i \dot{U}U^{\dagger}=H_{q+\Delta}&=H_{q}+\Delta U_q(t)\dot{J}(t)U_q^{\dagger}(t)+\mathcal{O}(\Delta^2)\\
       &=H_{q}+\Delta K(t)+\mathcal{O}(\Delta^2).\\
    \end{split}
   \end{equation}
   Where we have defined $K(t)=U_q(t)\dot{J}(t)U_q^{\dagger}(t)$.  To be sure that $H_{q+\Delta}$ is still a solution to the geodesic equation, we plug it into \eqref{eq:Schrodinger-geodesic}, and expand to first order in $\Delta$. To first order in $\Delta$, we find the constraint on $K$:
   \begin{equation}\label{kequation}
       \begin{split}
           0&=\dot{K}+i\mathcal{F}\Bigg(\left[K,\mathcal{G}(H_{q})\right]+\left[H_q,\mathcal{G}(K)\right]+\\
           &\mathcal{Q}(\mathcal{F}(\left[\mathcal{G}(H_q),H_q\right]))+\left[H_q,\mathcal{Q}(H_q)\right]\Bigg)\\
           &=\dot{K}+i\mathcal{F}\left(\left[K,\mathcal{G}(H_{q})\right]+\left[H_q,\mathcal{G}(K)\right]\right)+\\
           &i\mathcal{F}^2\left(\left[\mathcal{P}(H_q),\mathcal{Q}(H_q)\right]\right).
       \end{split}
   \end{equation}
    We used the fact that $\mathcal{P}\mathcal{Q} = 0$, $\mathcal{P}^2 = \mathcal{P}$,$\mathcal{Q}^2=\mathcal{Q}$.
   This is the so called "lifted" Jacobi equation. It is an inhomogeneous linear system, and so it's general solution can be written as:
   \begin{equation}
       K(t)=\mathcal{K}_t(K(0))-\mathcal{K}_t\left(\int_0^{t}d\tau \mathcal{K}_t^{-1}(C(\tau))\right)
   \end{equation}
   Where $\mathcal{K}_t$ is a linear operator which propagates the initial value of $K(0)$ such that it solves the homogenous part of the \eqref{kequation}, and $C=i\mathcal{F}^2\left(\left[\mathcal{P}(H_q),\mathcal{Q}(H_q)\right]\right)$ is the inhomogenous part of \eqref{kequation}. By expressing this equation in terms of coordinates, it turns out to be possible to derive an expression for $\mathcal{K}_t$. Expressing the homogeneous part of \eqref{kequation}, in terms of coordinates we find:
   
   \begin{equation}
   \begin{split}
       0&=\dot{K}_{ij}+i F_{ij}^{kl}\Big(K_{k\alpha}G_{\alpha l}^{\beta \delta}H^{(q)}_{\beta \delta}-G_{k \alpha}^{\beta \delta}H^{(q)}_{\beta \delta}K_{\alpha l}\\
       &+H^{(q)}_{k \alpha}G_{\alpha l}^{\beta \delta}K_{\beta \delta}-G_{k \alpha}^{\beta \delta}K_{\beta \delta}H^{(q)}_{\alpha l}\Big)\\
       &=\dot{K}_{ij}+i F_{ij}^{kl}\Big(\delta_{k\beta}G_{\delta l}^{\gamma \epsilon}H^{(q)}_{\gamma \epsilon}-G_{k \beta}^{\gamma \epsilon}H^{(q)}_{\gamma \epsilon}\delta_{\delta l}+\\
       &H^{(q)}_{k \alpha}G_{\alpha l}^{\beta \delta}-G_{k \alpha}^{\beta \delta}H^{(q)}_{\alpha l}\Big)K_{\beta \delta}\\
       &=\dot{K}_{ij}-i A_{ij}^{\beta \delta}K_{\beta \delta}.
    \end{split}
   \end{equation}
   Here we are using Einstein summation notation, and we have defined the tensor:
   
   \begin{widetext}
   \begin{equation}\label{AAAAAAAAAA}
       A_{ij}^{\beta \delta}=-F_{ij}^{kl}\Big(\delta_{k\beta}G_{\delta l}^{\gamma \epsilon}H^{(q)}_{\gamma \epsilon}-G_{k \beta}^{\gamma \epsilon}H^{(q)}_{\gamma \epsilon}\delta_{\delta l}+H^{(q)}_{k \alpha}G_{\alpha l}^{\beta \delta}-G_{k \alpha}^{\beta \delta}H^{(q)}_{\alpha l}\Big).
   \end{equation}
   \end{widetext}
   
    The homogeneous propagator can then simply be expressed as an ordered exponential:
   \begin{equation}\label{eq: k prop}
       \mathcal{K}(t)_{ijkl}=\mathcal{T}\left(\mathrm{e}^{i \int_0^t dt A(t)}\right)_{ijkl}
   \end{equation}
   Where we are suppressing indices inside of the time ordering operator for notational convenience. We can think of this as a normal matrix equation where the first two indicies of A correspond to the first index of a matrix and the last two indices of A correspond to the last index of a matrix. From this flattened perspective, A takes the form:
    \begin{align}
        \label{eq:A matrix}
        \nonumber A = F[(\mathbb{I} \otimes L - L^T \otimes \mathbb{I}) + (H^T \otimes \mathbb{I} - \mathbb{I} \otimes H)G] 
    \end{align}

    Note that in the equation above, we have replaced superoperators $\mathcal{F}$ and $\mathcal{G}$ with their matrix representations, 

\begin{align}
        F = \frac{1}{2^n}\sum _\beta \frac{1}{g(\beta)}\sigma^\beta \otimes \sigma^\beta \\
        \nonumber G = \frac{1}{2^n}\sum _\beta {g(\beta)}\sigma^\beta \otimes \sigma^\beta \\
        \nonumber g(\beta)= \begin{cases}
        1 & \text{if } \beta \leq 2, \\
        q & \text{else}.
        \end{cases}
\end{align}

This matrix representation is an eigen-decomposition of the relevant super-operators. We require that the lifted Jacobi field has boundary conditions $J(0) = J(T) = 0$, so the boundaries of the geodesic stay fixed. To accomplish this observe the relationship:
\begin{widetext}
\begin{equation}
\begin{split}
    J(T)&=\int_0^T dt\dot{J}(t)\\
    &=\int_0^T dt U^{\dagger}K(t)U\\
    &=\int_0^T dt U^{\dagger}\mathcal{K}_t(K(0))U-\int_0^T dt U^{\dagger}\left(-\mathcal{K}_t\left(\int_0^{t}d\tau \mathcal{K}_t^{-1}(C(\tau))\right)\right)U\\
    &=\mathcal{J}_T\left(\dot{J}(0)\right)-\int_0^T dt U^{\dagger}\left(-\mathcal{K}_t\left(\int_0^{t}d\tau \mathcal{K}_t^{-1}(C(\tau))\right)\right)U
    \end{split}
\end{equation}
\end{widetext}

Where we have identified $\dot{J}(0)=K(0)$. $J(0)$ is set to zero here as an initial condition for $K$. Because we need $J(T)=0$, we set it to zero in the above equation, and solve for $\dot{J}(0)$. The result is:
\begin{equation}\label{Jdot}
    \dot{J}(0)=\mathcal{J}_T^{-1}\left(\int_0^T dt U^{\dagger}\left(-\mathcal{K}_t\left(\int_0^{t}d\tau \mathcal{K}_t^{-1}(C(\tau))\right)\right)U\right).
\end{equation}
This gives us a way to update our initial Hamiltonian from q to $q+\Delta$ because according to \eqref{perturbed_ham}:
\begin{equation}
    \dot{J}(0)=\frac{\partial H_q(0)}{\partial q}\approx\frac{H_{q+\Delta}(0)-H_q(0)}{\Delta}.
\end{equation}
The right hand side (RHS) of \eqref{Jdot} simplifies and finally, a calculation shows:
\begin{widetext}
    \begin{equation}\label{eq:geodesic_derivative}
            \frac{dH_q(0)}{dq} =      \begin{cases}
        \mathcal{J}_T^{-1}(\intop_0^TdtU^\dag(t)it[\mathcal{P}(H), \mathcal{Q}(H)]U(t)), & \text{ q = 1,} \\
        (\mathcal{J}_T^{-1}(L(0))T - L(0))/q(q-1), & \text{ q > 1}, 
    \end{cases}
    \end{equation}
   \end{widetext}
   where $L(0)=\mathcal{G}(H(0))$. The base case for this update rule is given when $q=1$. In this case it is especially easy to calculate solutions to \eqref{eq:Schrodinger-geodesic} because the metric and its inverse are identity operators on the lie algebra. Concretely for this case we have:
       \begin{align}
    \begin{split}
        \label{eq:Schrodinger}
        \dot{U}\left(t\right)+\frac{i}{\hbar}H\left(t\right)U\left(t\right)&=0\\
        \dot{H}&=0,\\
        \end{split}
    \end{align}
   Which is still subject to the boundary condition: $U(0)=\mathbb{1}$, and $U(T)=U_{\mathrm{T}}$. In this case, the Hamiltonian is a constant and its initial value is given by $H_1(0)=\frac{i}{T}\mathrm{log}\left(U_T\right)$. A derivation of this solution will be presented in the analytical section. Given this initial Hamiltonian and the update rule given at \eqref{eq:geodesic_derivative} it is possible to construct solutions that hit the target unitary for arbitrarily large penalty factors.

\section{Analytic Approximations} \label{sec:AA}
    In this section, we consider analytic solutions and approximations to our problem. Analytic solutions and approximations give context for our full numerical solutions. In the 1 and 2 qubit cases, $\mathfrak{su}\left(2\right)$ and $\mathfrak{su}\left(4\right)$, the geodesic equation simplifies to $\dot{\hat{H}}=0$ because $\mathcal{G}$ is the identity operator on traceless 2 qubit operations. This gives us analytic values on the complexities which we can then compare to our numerical solutions.

    When $\dot{\hat{H}}=0$, we have the solution on the interval $\left[0,T\right]$
    
    \begin{equation}
        U(t)={\mathrm e}^{-i \hat{H} t}U_{0},
    \end{equation}
    
    Where we are imposing the boundary conditions: $U(0)=\mathbb{I}$, and $U(T)=U_{\mathrm{T}}$.  The boundary condition at time $T$ fixes the Hamiltonian in this case:
    
    \begin{equation}
        \hat{H}=\frac{i}{T}\mathrm{log}\left(U_T\right)
    \end{equation}
    Here we are computing logarithms of matrices by first diagonalizing the matrix and taking the logarithm of the eigenvalues. Because the $U_T$ is a unitary matrix, the eigenvalues are complex phases. When we take logarithms, we compute the phase up to modulo $\pi$. This uniquely determines the Hamiltonian despite the ambiguity associated with the logarithm.
    From here, plugging into \eqref{metric} and \eqref{complexity}, we find the approximate complexity to be
    
    \begin{equation}
        C\approx\sqrt{-\frac{1}{2^{n}}\mathrm{tr}\left(\mathrm{log}\left(U_T\right)\mathcal{G}\left(\mathrm{log}\left(U_T\right)\right)\right)}
    \end{equation}
    This is the complexity associated with a straight line path from the identity operator to the target unitary. For one and two qubits it is an exact value of the complexity and for high numbers of qubits it serves as an approximation to which we can  compare our exact answers.

\subsubsection{One Qubit}
In the case of one qubit, the lie algebra is $su(2)$, with three dimensions spanned by $\sigma_x, \sigma_y $ and $\sigma_z$. General elements in this algebra are given by $H=V^i\sigma_i$, and the inner product simplifies to:
\begin{equation}
    \begin{split}
        \langle H,J\rangle&=\frac{1}{2}tr\left(H \mathcal{G}(J)\right)\\
        &=H^iJ^j\frac{1}{2}tr(\sigma_i \sigma_j)\\
        &=g_{ij}H^iJ^j.
    \end{split}
\end{equation}
Where we have defined
\begin{align}
    \label{eq:Metric n=1}
    g_{ij}=\left[\begin{array}{ccc}
    1 & 0 & 0\\
    0 & 1 & 0\\
    0 & 0 & 1
    \end{array}\right]
\end{align}
This is the normal inner product one uses on Euclidean space. In the following, we consider the target unitary matrix, which corresponds to the quantum Fourier transform on one qubit:

\begin{align}
    U_{T}=\frac{-i}{\sqrt{2}}\left[\begin{array}{cc}
    1 & 1\\
    1 & -1
    \end{array}\right]
\end{align}
To construct a geodesic that hits this target, we first compute the matrix logarithm to obtain the Hamiltonian. This is most easily done using computer algebra and the result is:
\begin{align}
    H=\frac{\pi}{2\sqrt{2}}\left[\begin{array}{cc}
    1 & 1\\
    1 & -1
    \end{array}\right]
\end{align}

Now to obtain the geodesic connecting the identity and target, we need to solve the Shr\"odinger equation: $\dot{U}=-i H U$. Because the Hamiltonian is a constant here, this can be done by using the matrix exponential: $U=\mathrm{e}^{-i H t}$. Computing this quantity and simplifying, we find:

\begin{align}
    U\left(t\right)=\left[\begin{array}{cc}
    \alpha\left(t\right)-\beta\left(t\right) & -\beta\left(t\right)\\
    -\beta\left(t\right) & \alpha\left(t\right)+\beta\left(t\right)
    \end{array}\right]\\ \nonumber
    \alpha\left(t\right)=\cos\left(\frac{\pi}{2}t\right)\ \ \ \beta\left(t\right)=\frac{i}{\sqrt{2}}\sin\left(\frac{\pi}{2}t\right)
\end{align}
Finally, it is possible to compute the complexity. It is simply given by $C=\sqrt{\frac{1}{2}tr(HH)}=\frac{\pi}{2 \sqrt{2}}$.
\subsubsection{Two Qubits}

In the case of two qubits, the lie algebra we work with is $su(4)$ and it spanned by Kronecker products of at most two pauli matrices. This space is 15 dimensional and the different basis elements are shown in  \ref{eq:Metric Basis n=2}, and because no three qubit operations are allowed at this stage, the super-operator $\mathcal{G}$ is still an identity operation on traceless unitary matrices. Because of this, the metric tensor is given by:

\begin{equation}
    g_{ij}=\delta_{ij}
\end{equation}
Where $\delta_{ij}$ is the normal Kronecker delta, corresponding to a Euclidean inner product on this 15 dimensional space.


We consider now the following target unitary matrix, which corresponds to the two qubit quantum fourier transform:

\begin{align}
    \label{eq:Target Unitary n=2}
    U_{T}=\frac{1}{\sqrt{2^{n}}}\left[\begin{array}{cccc}
    1 & 1 & 1 & 1\\
    1 & \omega & \omega^{2} & \omega^{3}\\
    1 & \omega^{2} & 1 & \omega^{2}\\
    1 & \omega^{3} & \omega^{2} & 1
    \end{array}\right]
\end{align}

Where we have defined the variable:
\begin{align}
    \label{eq:omega n=2}
    \omega=e^{\pi i/2}.
\end{align}

As in the previous section, we then compute the Hamiltonian using matrix logarithms, and use that Hamtilonian in a matrix exponential to construct the geodesic path to the target unitary matrix. Up to a global phase, we find that path is given by:

\begin{align}
    \label{eq: n=2 Unitary}
    U\left(t\right)=\frac{1}{4}\omega^{-\frac{3}{4}t}\left[\begin{array}{cccc}
    \alpha & \gamma & \gamma & \gamma\\
    \gamma & \beta & \delta & \epsilon\\
    \gamma & \delta & \alpha & \delta\\
    \gamma & \epsilon & \delta & \beta
    \end{array}\right]\\ \nonumber
    \alpha=\omega^{2t}+3,\ \ \ \beta=\left(1+\omega^{t}\right)^{2}, \ \ \
    \gamma=1-\omega^{2t},\\ \nonumber
    \delta=\omega^{2t}-1, \ \ \ \epsilon=\left(\omega^{t}-1\right)^{2}
\end{align}

Again, we can compute the complexity according to $C=\sqrt{\frac{1}{4}tr(HH)}=\frac{\sqrt{11}\pi}{16}$.

\subsubsection{Several Qubits}

When we have more than two qubits, the solution to the geodesic equation is no longer  a constant Hamiltonian; this makes the unitary evolution determined by Schr{\"o}dinger's equation \eqref{eq:Schrodinger} much more difficult to calculate analytically. As a first approximation, in this section, we consider paths in the unitary group corresponding to constant Hamtilonians which hit the target. These paths are not geodesic and so do not minimize the complexity, but they do put upper bounds on the complexity which can be used to give our numerical solutions context. 

Our procedure for doing this is the same as the two previous subsections. Starting with the target unitary matrix, we first compute the Hamiltonian using matrix logarithms. After that, the path in the unitary group is found by computing the Matrix exponential: $U=\mathrm{e}^{-i H t}$. Finally, the complexity is computed according to the formula $C\approx\langle H,H\rangle ^{1/2}$. We have computed this approximate complexity for the quantum fourier transform and for the CNOT gate for up to nine qubits. The results are shown in the figure below which also contains numerically computed geodesic complexities for up to four qubits.

\begin{figure}[ht]
    \centering
    \includegraphics[width=\linewidth]{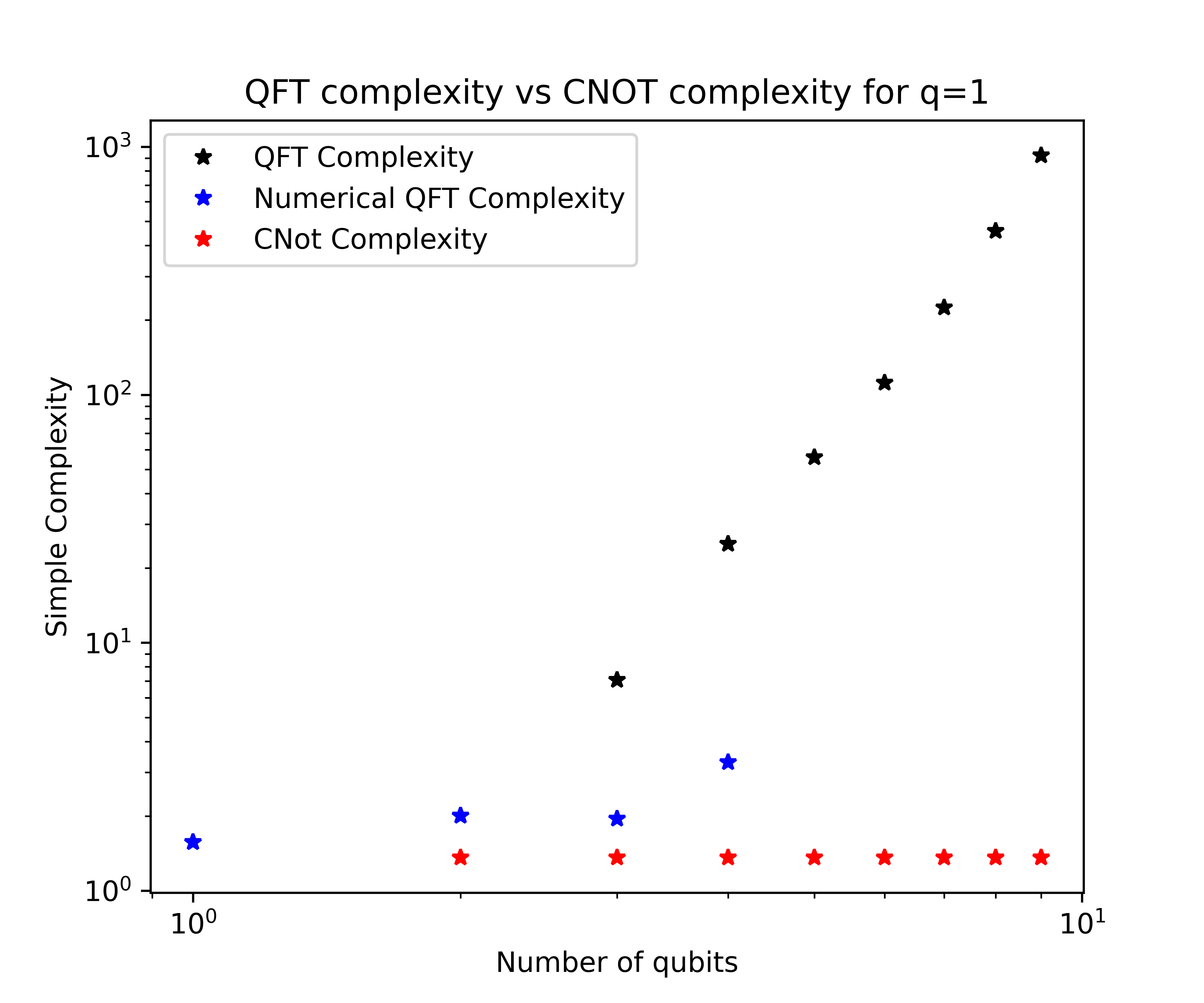}
    \caption{Comparison of simple complexity, Numerical complexity, ($q = 4^n-1$) case, between quantum Fourier transform and CNOT operator as number of qubits varies.}
    \label{fig:simple_comp}
\end{figure}



\section{Numerical Analysis} \label{sec:NA}
    In this section we discuss the subtleties associated with creating a numerical program that can compute quantum computational geodesics.

    To make unitary matrices that correspond to quantum circuits we use a package in \url{qiskit} that has built in functionality to do this. For example, constructing the $N$ qubit quantum Fourier transform matrix representation can be accomplished with the commands:

    \begin{lstlisting}
from qiskit.quantum_info import Operator
from qiskit.circuit.library import QFT
#Get the QFT, object.
operator = Operator(QFT(N_qubits))
# Get the matrix representation
U_target = operator.data
\end{lstlisting}

This procedure is quite general and can be used to construct the matrix representation of any qiskit circuit object. Once the target unitary operator has been obtained we then compute the Hamiltonian $H_0$ for the desired unitary at time $T = 1$ in the case where $q = 1$, taking a matrix logarithm as suggested by \eqref{eq:Schrodinger Sol w const H}. To compute matrix logarithms, we developed our own code to have tighter controls over ambiguities associated with logarithms. In this way, we compute logarithms using the code:

\begin{lstlisting}
@njit
def complex_unitary_log(
    U:np.ndarray
) -> np.ndarray:
    
    eigvals, eigvecs = np.linalg.eig(U)
    eigvecs_inv = np.linalg.inv(eigvecs)
    eigvals = np.exp(1j*np.angle(eigvals))  # np.angle shifts phase to (-pi, pi)
    log_diag = np.diag(np.log(eigvals))

    return 1j * eigvecs @ log_diag @ eigvecs_inv
\end{lstlisting}
This logarithm works by first diagonlizing the unitary matrix and then taking the log of the eigenvalues modulo $\pi$. The basis is then changed to the original one and the log of the matrix is returned.

Generally, all functions we use are decorated with the \url{@njit} operator such that functions are precompiled and run much faster at runtime.

Because we are only considering target unitaries that can be generated by elements of the lie algebra $su(2^n)$, we substract the zero point energy of the Hamiltonian. This amounts to a constant shift of the energy levels of the Hamiltonian which does not affect its dynamics. This is accomplished by: $H' = H_0 - \frac{1}{2^n}tr(H_0)\mathbb{I} $, which effectively makes the Hamiltonian traceless. 

At this stage we need to compute the RHS of \ref{eq:geodesic_derivative}. To begin we develop a numerical implementation of the super operators: $\mathcal{F}$ and, $ \mathcal{G}$. This can be done by implementing the super operators $\mathcal{P}$, and $\mathcal{Q}$ numerically. Generally, the operators  $\mathcal{P}$, and $\mathcal{Q}$ operate on are of the form $V^i \sigma_i$, where $V^i$ are real coefficients and $\sigma_i$ is a Pauli string. For three qubits all Pauli strings are shown in \ref{eq:Metric Basis n=3}. From there the coefficient can be obtained from $V^i=\frac{1}{2^n}tr(V^j\sigma_j \sigma_i)$, one can then count the number of pauli matrices in a given Pauli string which is associated with $V^i$. For example, $\mathcal{P}(V^i\sigma_i)=\sum_i w_i V^i \sigma_i $, where $w_i$ goes to zero if the weight is greater than 2 and one otherwise. In python code this is implemented by:
\begin{lstlisting}
@njit
def get_coeff_and_basis(
    a:np.ndarray,
    pauli_signature:tuple | str
) -> Tuple[complex, np.ndarray]:
    
    pauli_dict = get_pauli_dict()
    basis_matrix = basis_constructor([pauli_dict[key] for key in pauli_signature])
    basis_coeffi = pauli_inner_product(a, basis_matrix)

    return basis_coeffi, basis_matrix

@njit
def PQ_decomp(
    a:np.ndarray,
    /
) -> np.ndarray:
    
    num_rows = a.shape[0]
    n_qubits = 0
    while num_rows > 1:
        num_rows >>= 1
        n_qubits += 1
    
    p = np.zeros_like(a)

    for signature in generate_P_signatures(n_qubits=n_qubits):

        coeff, basis_matrix = get_coeff_and_basis(a, pauli_signature=signature)
        p += coeff*basis_matrix

    return p, a - p
\end{lstlisting}
    Once the super operators $\mathcal{P}$, and $\mathcal{Q}$ have been implemented we need to solve for the time evolution of the Hamiltonian and $U(t)$, essentially, numerically finding solutions to \ref{eq:Schrodinger-geodesic}. We do this using the \url{scipy.integrate.solve_ivp} function. We solve over the range $t\in \left[0,1\right]$, and we record the solution value at $N_t$ linearly spaced points along this interval. These points will be used later on for computing the integrals in \eqref{eq:geodesic_derivative}. During the solution however, solve\_ivp may take as many steps as it needs to satisfy its default error tolerance.

    Numerically now, we must calculate the Jacobi propagator. Again, this object is defined as
    \begin{equation}
        \mathcal{J}_T\left(\dot{J}(0)\right)=\int_0^T dt U^{\dagger}\mathcal{K}_t(\dot{J}(0))U.    
    \end{equation}
    Expressing this in terms of coordinates we find:
    \begin{equation}
        \mathcal{J}_T^{ij \gamma \delta}\dot{J}_{\gamma \delta}(0)=\left(\int_0^T dt U^{\dagger}_{i \alpha}\mathcal{K}^{\alpha \beta \gamma \delta}_tU_{\beta j}\right)\dot{J}_{\gamma \delta}(0). 
    \end{equation}
    The K propagator can be computed according to \ref{eq: k prop}. We find it is most efficient to compute the ordered matrix exponential using a product: 
    \begin{equation}
        \mathcal{K}^{\alpha \beta \gamma \delta}_t=\prod_{i=1}^{i=t_f}\mathrm{e}^{i \Delta t_i A(t_i)}.
    \end{equation}
    Where values for $A$ are determined from our solutions for $H$, and $U$, which have been recorded on $N_t$ linearly spaced points on $\left[0,1\right]$. Concretely, $A$ is calculated from \ref{AAAAAAAAAA}.  It is understood that operators evaluated at later times are placed towards the left. Putting all of this together, we then compute $\mathcal{J}_T^{ij \gamma \delta}=\left(\int_0^T dt U^{\dagger}_{i \alpha}\mathcal{K}^{\alpha \beta \gamma \delta}_tU_{\beta j}\right)$. In python, this can be done with one line:
    \begin{lstlisting}
J_prop=np.trapz(np.einsum('iad,idekl,ieb->iaklb',U_dagger_o_t,k_tensor,U_o_t),time,axis=0)
\end{lstlisting}
To obtain the inverse Jacobi propagator we reshape this tensor into a matrix and then invert it. Using this technique, we are able to evaluate \eqref{eq:geodesic_derivative} and we use solve\_ivp to solve it as a differential equation. In the following section, we discuss results.

\subsection{Results} \label{sec:RD}
     

In this section we consider two numerical examples, and discuss our results. As a first numerical example, we consider the three  and four qubit Quantum Fourier transform. As a second example, we consider a randomly generated quantum circuit with a depth of 100 gates.  The randomly generated circuit can be seen in Figure~\ref{fig:random_circuit}. The three and four qubit quantum fourier transform can be seen in Figure~\ref{fig:3qft} and Figure~\ref{fig:4qft}

\begin{figure}[ht]
    \centering
    \includegraphics[width=\linewidth]{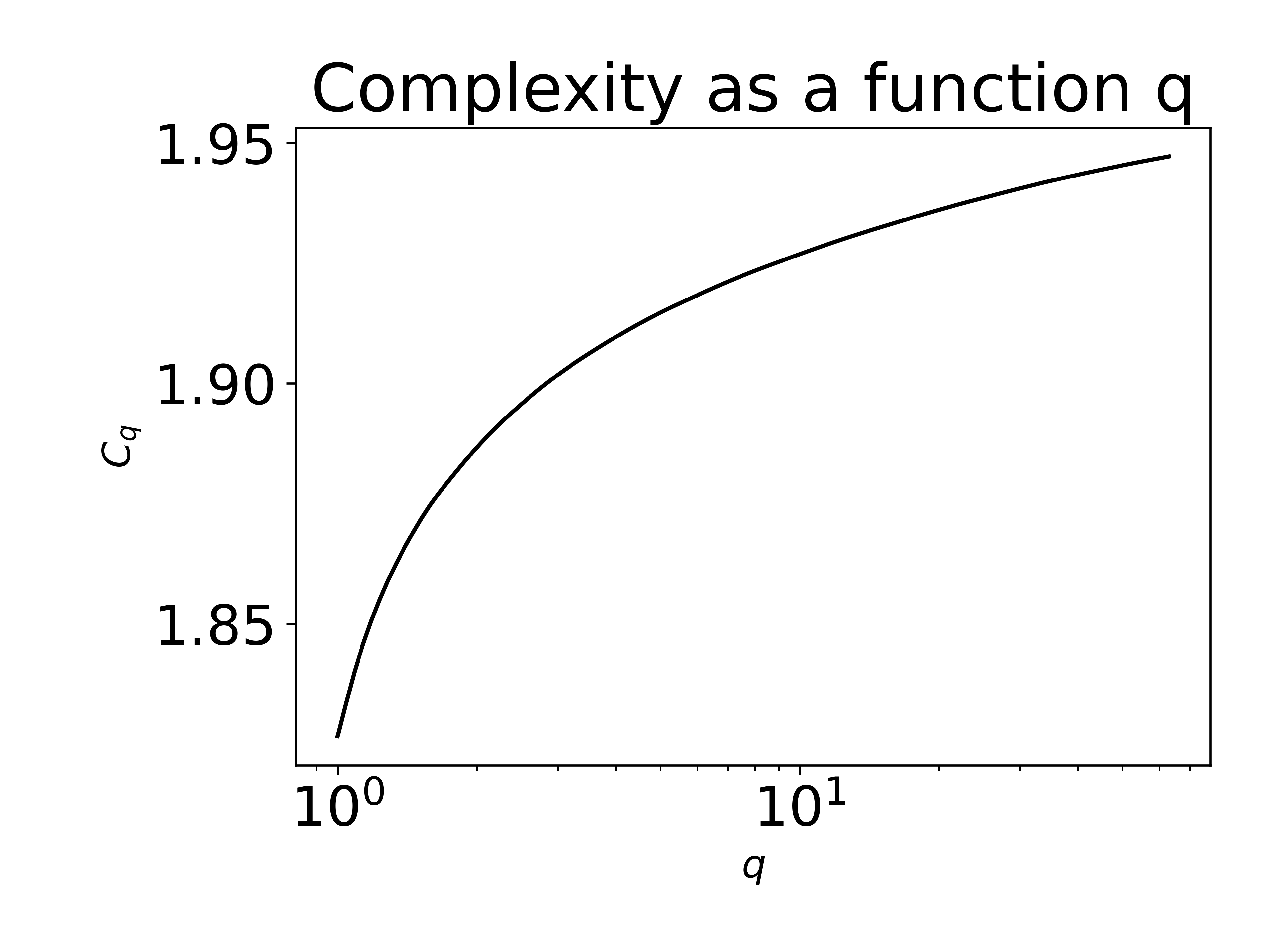}
    \caption{The complexity of the three qubit QFT as a function of the penalty factor}
    \label{fig:QFT_complexity_v_q}
\end{figure}

Following the procedure outlined in the previous sections, we compute the complexity of the three qubit quantum Fourier transform. The complexity as a function of the penalty parameter can be seen in Figure~\ref{fig:QFT_complexity_v_q}. As the penalty parameter is increased the complexity increases also but reaches an asymptote close to $C=1.95$. We have computed the complexity of a CNOT gate using our procedure and it has a complexity of $C_{\mathrm{CNOT}}=\frac{\sqrt{3}\pi}{4}\approx 1.4$. The three qubit quantum fourier transform can therefore be implemented with $1.4$ CNOT gates worth of effort. The standard quantum circuit implementation requires two control phase gates and a Hadamard gate. 

\begin{figure*}[ht]
    \centering
    \includegraphics[width=\textwidth]{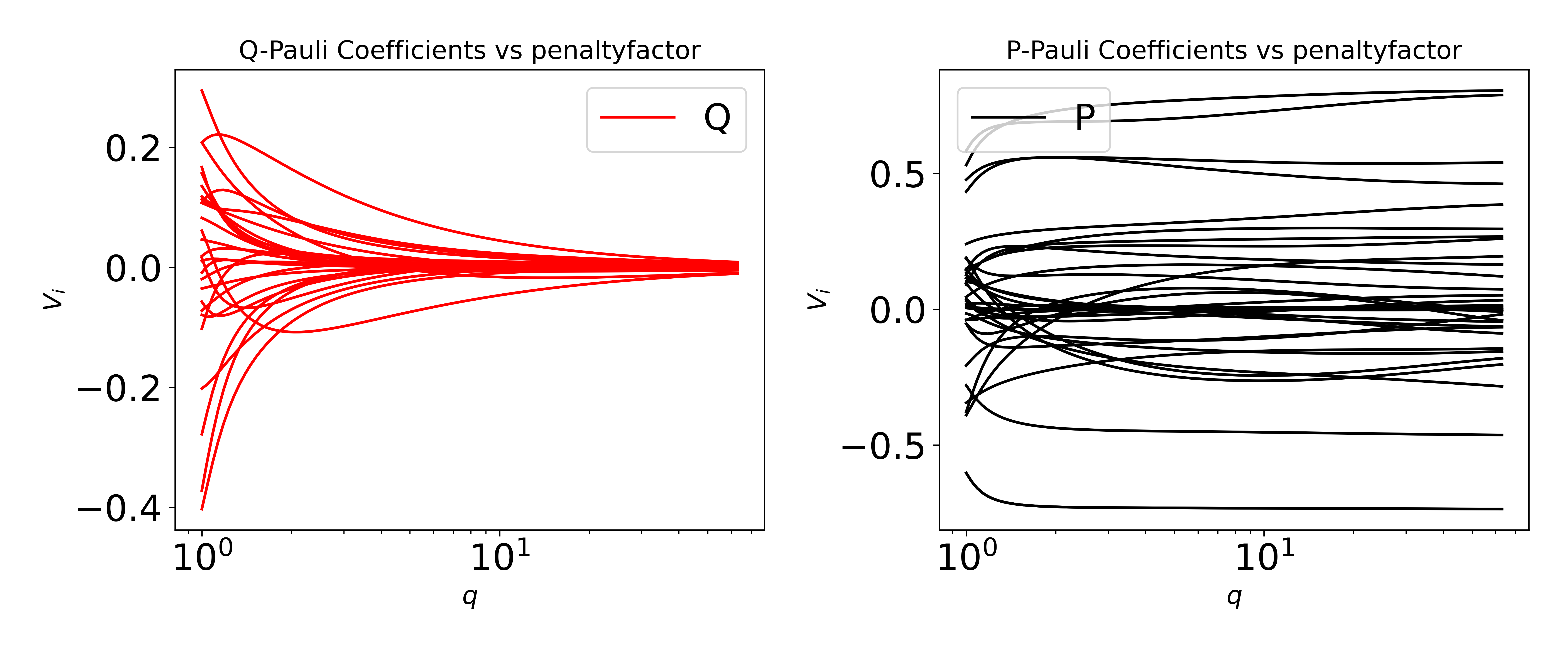}
    \caption{Initial coefficients of the Hamiltonian as a function of q. These coefficients are such that under geodesic evolution the target unitary will be hit.}
    \label{fig:QFT_coeffs_v_q}
\end{figure*}

The initial Pauli coefficients for our Hamiltonian as a function of the penalty factor can be seen in Figure~\ref{fig:QFT_coeffs_v_q}. In this figure, Pauli strings with weight higher than two are shown in red and labeled ``Q'' and these are the generators who are penalized for being difficult to implement in the lab. We notice that as the penalty factor increases these penalized generators have shrinking amplitudes indicating that they are in fact being penalized by the algorithm. Moreover we note that the initial coefficients seem to be reaching steady state values indicating our cutoff value for q is a reasonable one. 

After having found the correct initial coefficients for our Hamiltonian, the geodesic equation can then be solved to obtain the time dependent Hamiltonian coefficients. For this particular example the time dependent coefficients can be found below in Figure~\ref{fig:QFT_coeffs_v_t}. For concrete implementations of these coefficients on a quantum computer one would use these as modulation functions for a carrier wave which would then be directed towards the qubits. Geometrically these coefficients are interpreted as tangent vectors, from quantum control perspective they are interpreted as control functions. A physical realization of such a procedure is beyond the scope of this report and we reserve it for later.

\begin{figure}[ht]
    \centering
    \includegraphics[width=\linewidth]{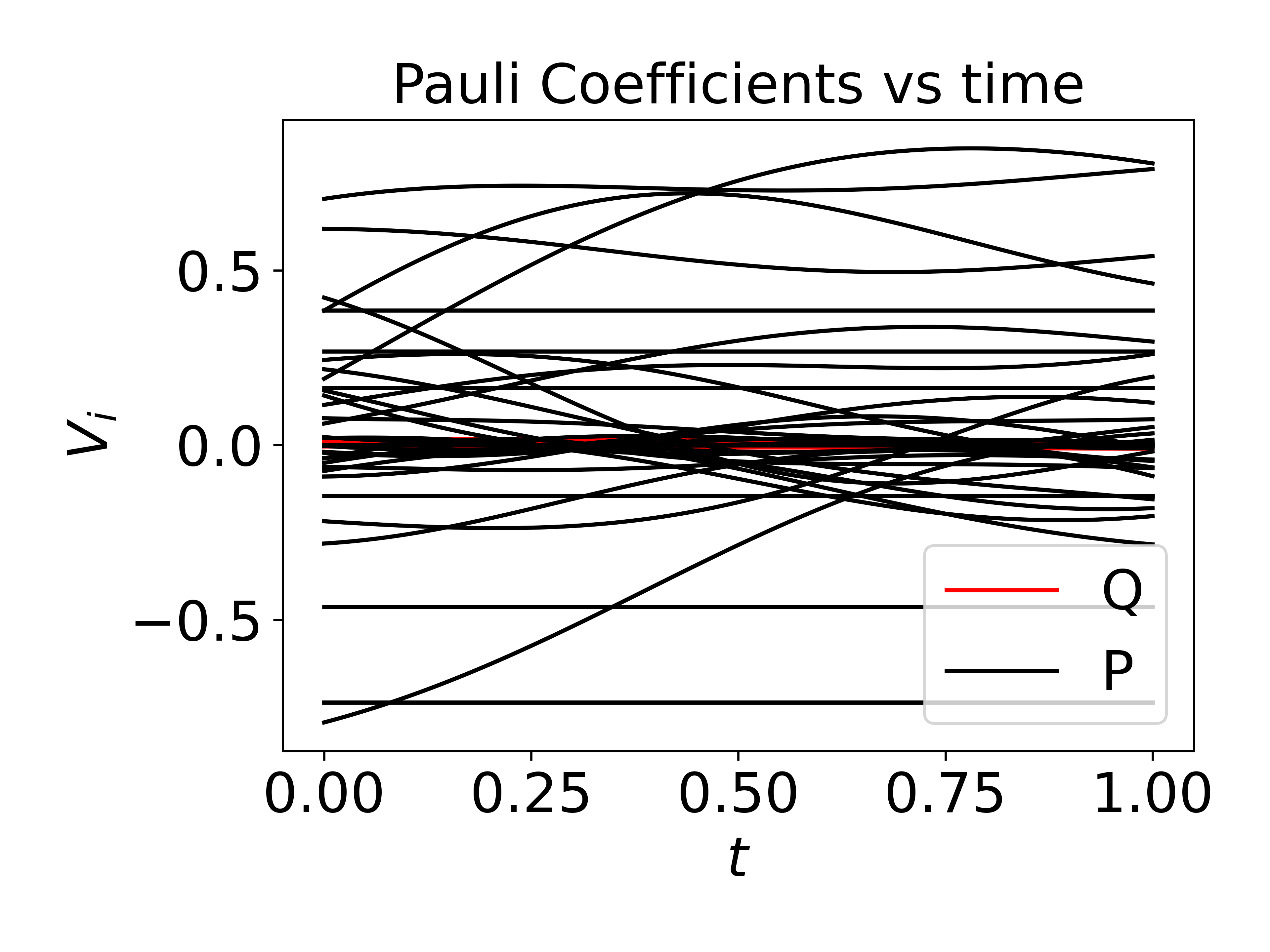}
    \caption{Pauli coefficients of the Hamiltonian vs. time for large penalty factor.}
    \label{fig:QFT_coeffs_v_t}
\end{figure}

A version of this plot where only some control functions are shown and labeled is available in the appendix in \ref{fig:labled_control}.

We have also repeated this analysis for the four qubit quantum fourier transform which is shown in \ref{fig:4qft}. We find the complexity to be $3.29$ and the control functions are shown in Figure~\ref{fig:QFT_coeffs_4_v_t}. A selection of control functions with labels can be found in the appendix for the four qubit quantum fourier transform.

\begin{figure}[h]
    \centering
    \includegraphics[width=\linewidth]{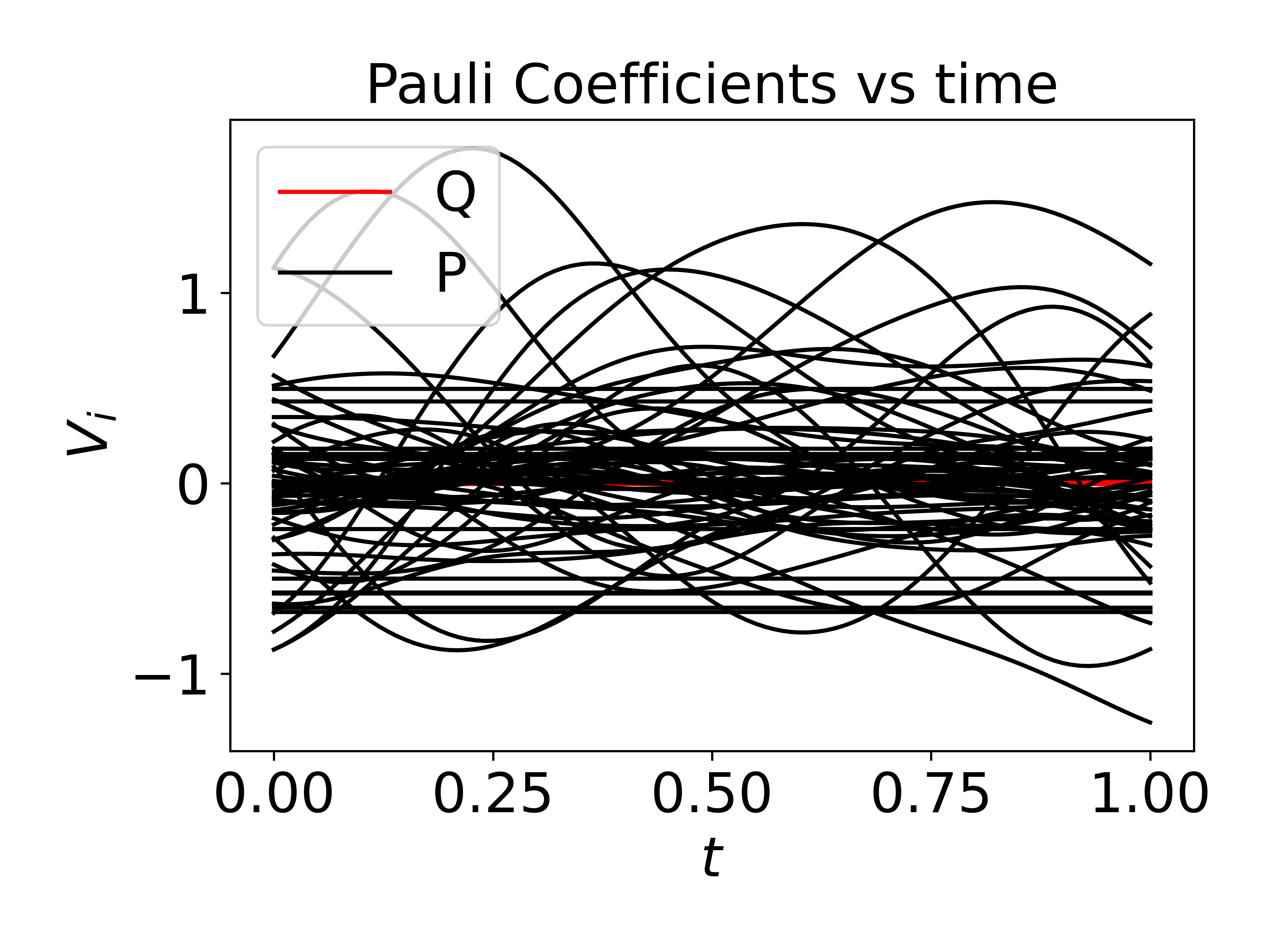}
    \caption{Pauli coefficients of the Hamiltonian vs. time for large penalty factor in the case of the four qubit quantum fourier transform.}
    \label{fig:QFT_coeffs_4_v_t}
\end{figure}

As a final step, we verify that the geodesic evolution of our unitary matrix indeed hits the target unitary. This can be checked by computing the rms deviation of our time dependent unitary with the target unitary. That is we compute $||U(t)-U_{\mathrm{Target}}||=\sqrt{\frac{1}{4^n}\left(U^{\dagger}(t)-U^{\dagger}_{\mathrm{Target}}\right)\left(U(t)-U_{\mathrm{Target}}\right)}$ for $t\in \left[0,1\right]$. Finally, we then plot this error function as a function of time. This plot can be seen below in Figure~\ref{fig:QFT_dU_v_t}. We notice that the rms deviation approaches zero as time goes to $1$, indicating that we are indeed hitting the target.

\begin{figure}[ht]
    \centering
    \includegraphics[width=\linewidth]{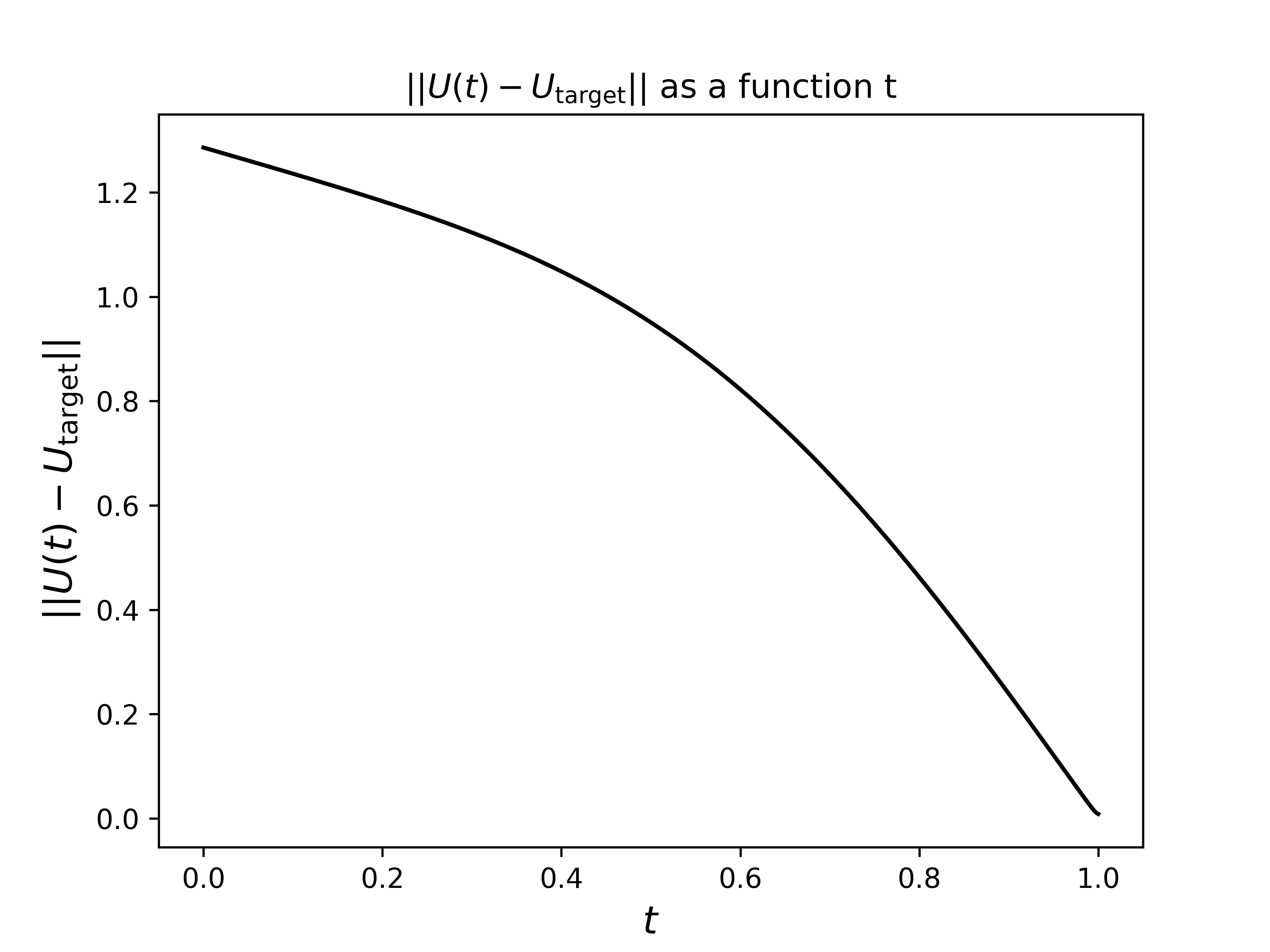}
    \caption{Verifying that the $U(t)$ numerical solution evolves to $U(T) = U(1) = U_{target}$.}
    \label{fig:QFT_dU_v_t}
\end{figure}

As a second example, we compute the complexity of a randomly generated quantum circuit with a depth of 100, which can be seen in Figure~\ref{fig:random_circuit}. Our computation of the complexity of this quantum circuit as a function of the penalty factor can be seen below in Figure~\ref{fig:RAND_complexity_v_q}. We notice that the complexity is increasing with increasing penalty factor and reaches an asymptote close to $C=2.4$. So despite the perceived complexity of the circuit shown in Figure~\ref{fig:RAND_complexity_v_q}, it can actually be executed with a complexity of only $2.4$ corresponding to $1.8$ CNOT gates worth of effort. A capability of the QGeo python package is the ability to uncover such speed ups.

\begin{figure}[ht]
    \centering
    \includegraphics[width=\linewidth]{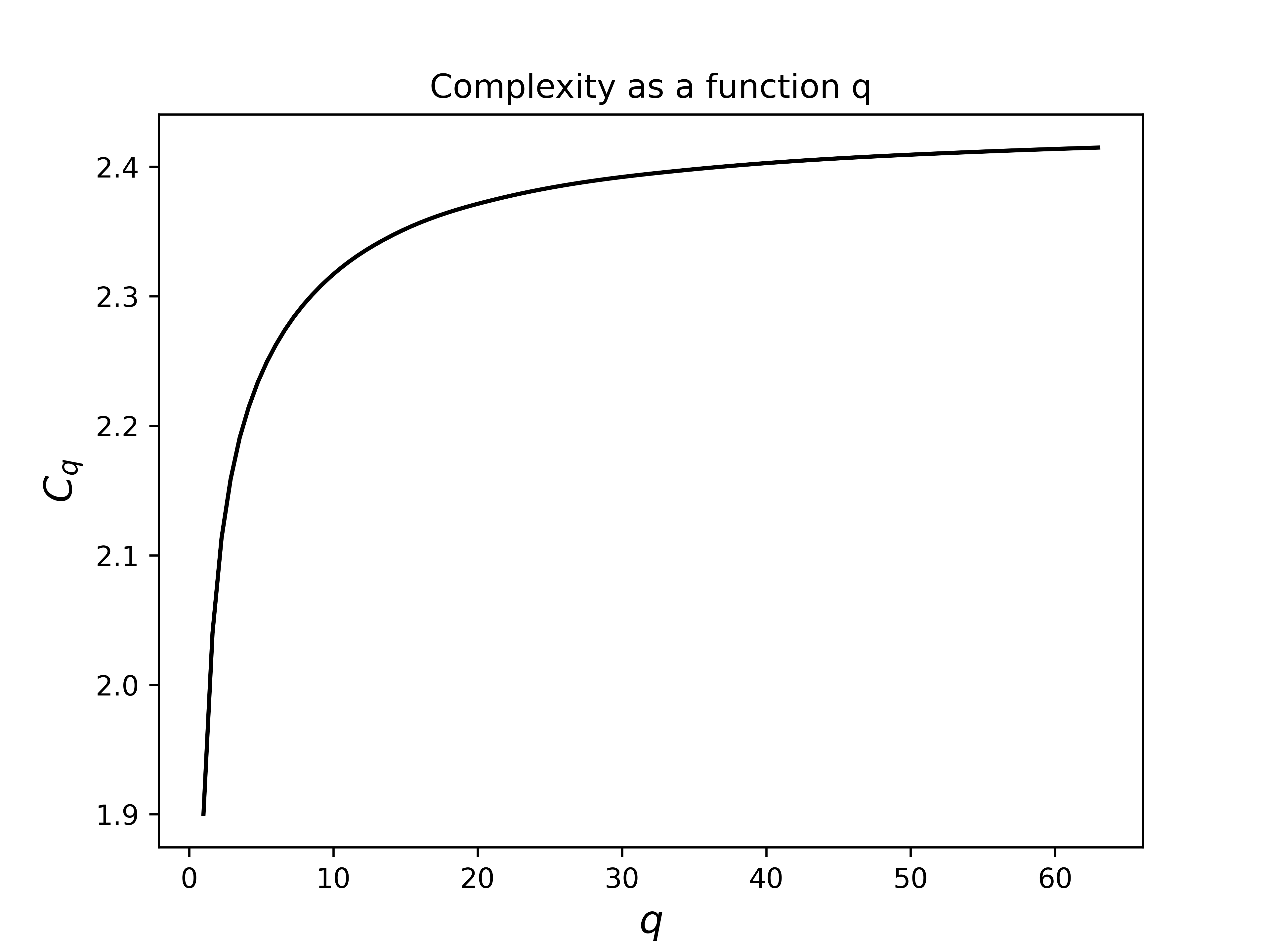}
    \caption{Initial coefficients of the Hamiltonian as a function q for the randomly generated quantum circuit}
    \label{fig:RAND_complexity_v_q}
\end{figure}

\begin{figure}[ht]
    \centering
    \includegraphics[width=\linewidth]{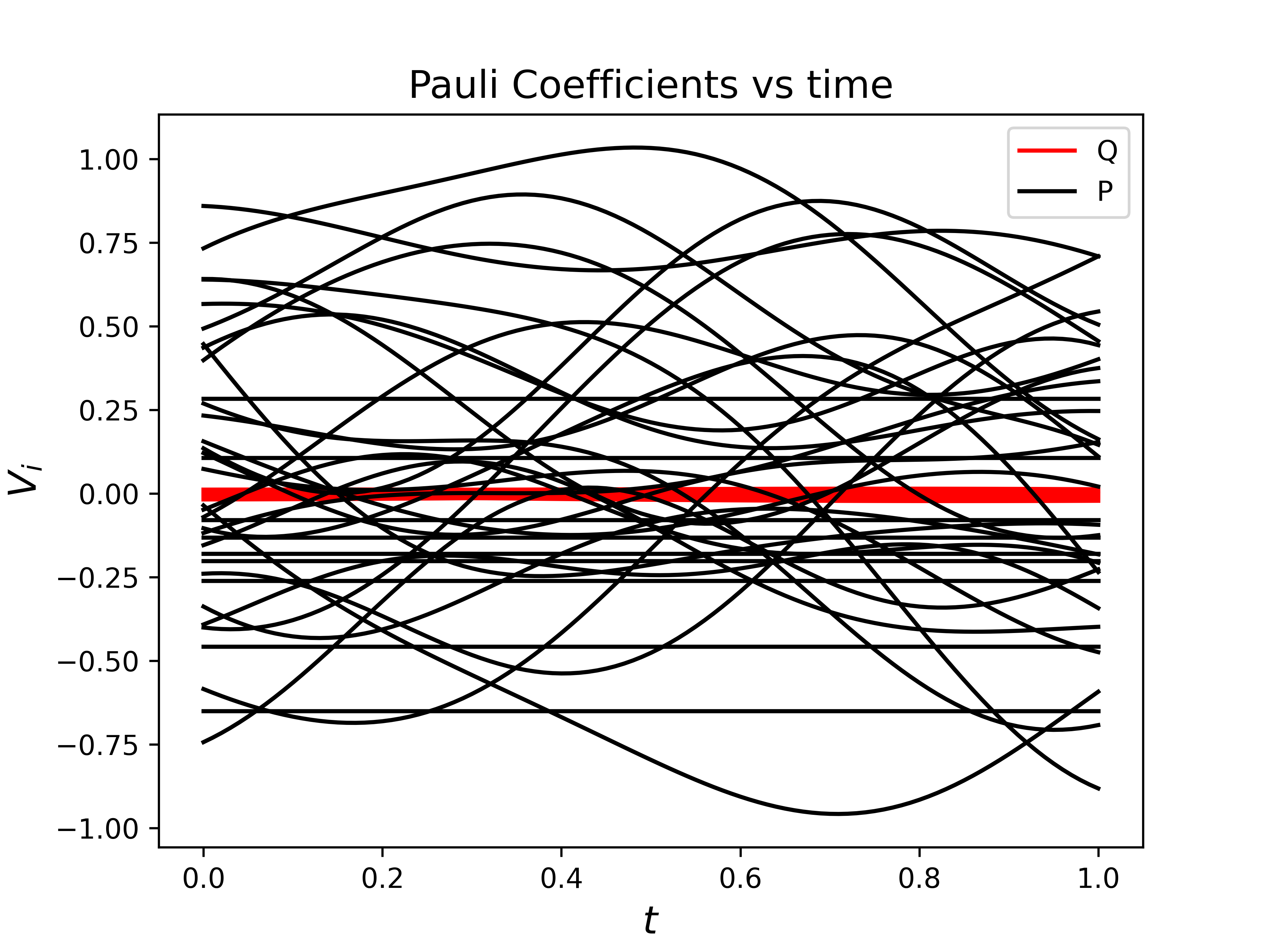}
    \caption{Time evolution of the Hamiltonian coefficients for the randomly generated circuit. }
    \label{fig:RAND_coeffs_v_t}
\end{figure}

After having determined the initial values of the Hamiltonian for this randomly generated circuit, we then time evolve the Hamiltonian to obtain its time evolution. The time evolution of the Hamiltonian coefficients can be seen in Figure~\ref{fig:RAND_coeffs_v_t}. These coefficients are considerably more complicated looking than the ones found in the case of the quantum fourier transform, however the amplitude here is lower, so the resulting complexity is still about the same.

Finally, we compute the RMS deviation of our time dependent unitary with the target unitary. The time dependent RMS deviation can be seen below in Figure~\ref{fig:RAND_dU_v_t}. Again, we are able to hit the target.

\begin{figure}[ht]
    \centering
    \includegraphics[width=\linewidth]{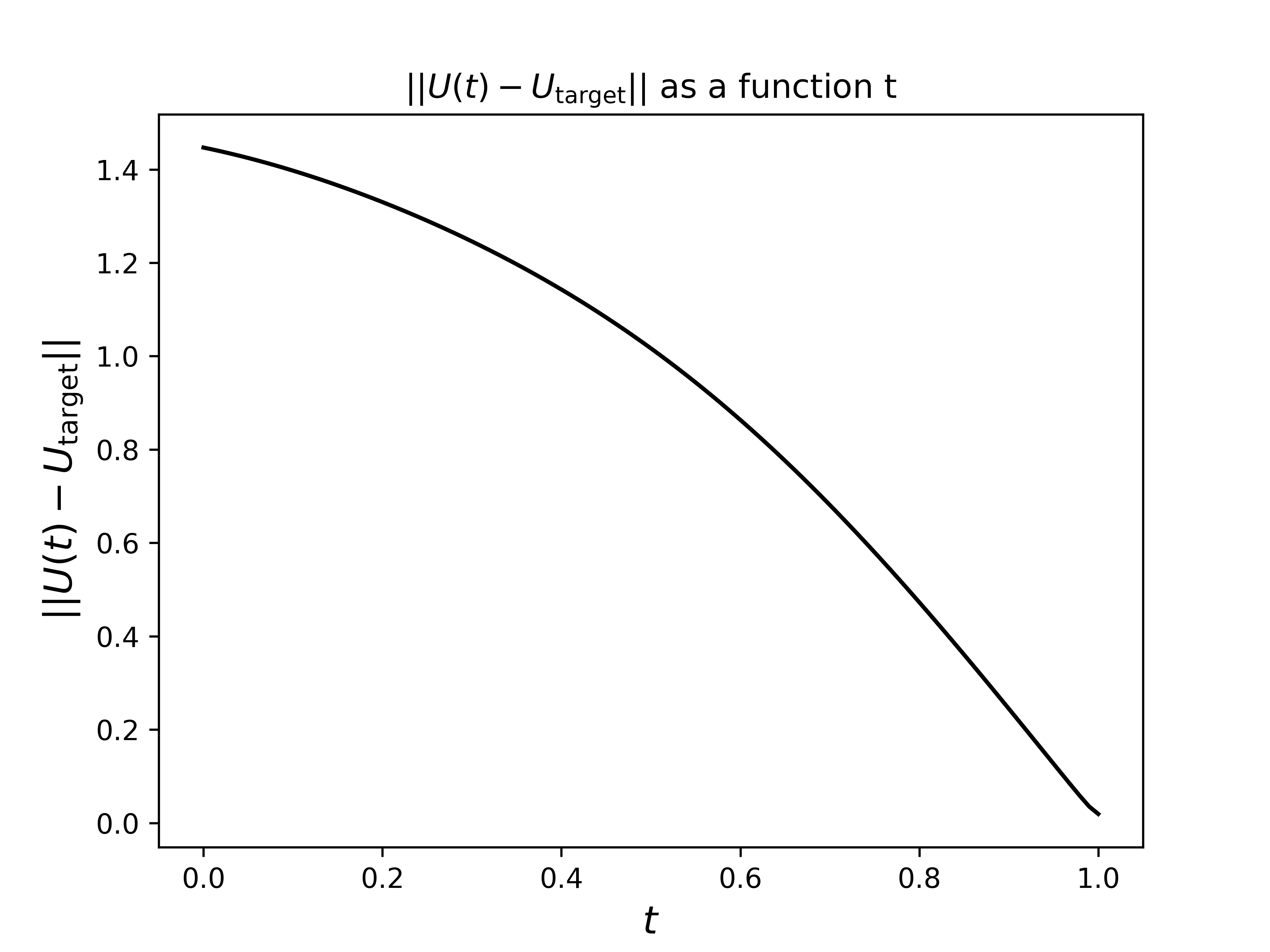}
    \caption{Verifying that the $U(t)$ numerical solution evolves to $U(T) = U(1) = U_{target}$.}
    \label{fig:RAND_dU_v_t}
\end{figure}

\subsection{Fermion Chain Evolution}
As a final example of our implementation of this algorithm, we calculate the quantum complexity of parts of a quantum circuit which is used to calculate the evolution of a Hubbard model on a fermion chain. Our method can only process a small number of qubits.  However, we can use our method to reduce the depth of a large circuit if the circuit can be built by combining smaller circuits.   As an example, let us consider a fermionic chain. In the case of non-interacting fermions, this chain is described by the Hamiltonian:

\begin{figure}[ht]
\centering
\includegraphics[width=\linewidth]{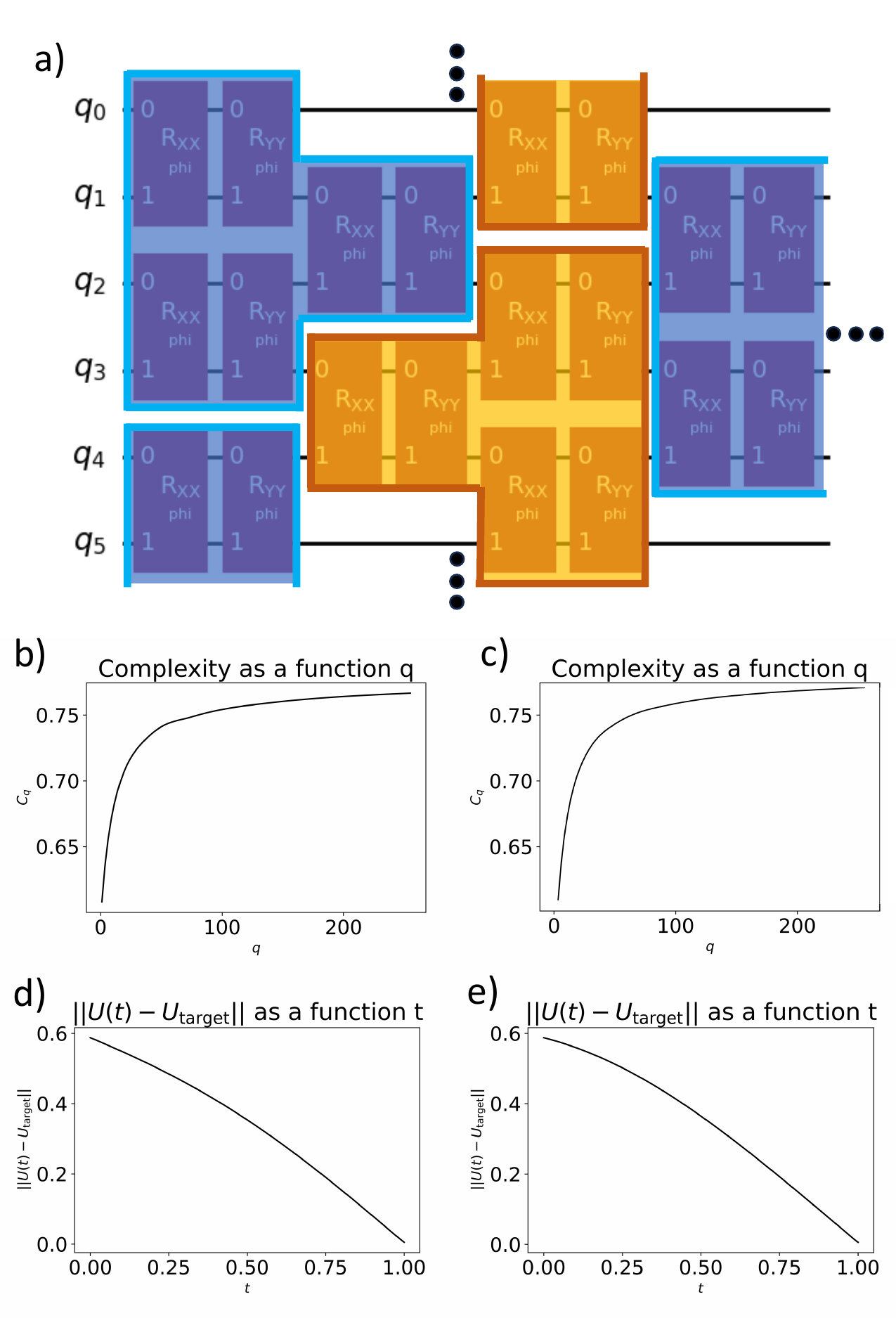}
\caption{a) trotterized fermionic evolution circuit.  The circuit is divided into two types of blocks.  Block 1 is shaded blue and block 2 is shaded orange.  b) complexity of the block 1 circuit as a function of q.  c) complexity of the block 2 circuit as a function of q.  d) Error of the block 1 circuit as a function of time.  Error of the block 2 circuit as a function of time.}
\label{fblock}
\end{figure}

\begin{equation}
    H_{\text{Fermion}} = h\sum_{i} \left(c^{\dagger}_i c_{i+1} + c^{\dagger}_{i+1} c_i \right).
\end{equation}
Using the Jordan-Wigner transformation it is possible to express this Hamiltonian in terms of Pauli matrices acting at different sites, allowing for an implementation on a quantum computer. This transformation is given by:
\begin{equation}
\begin{split}
    c^{\dagger}_j = \frac{1}{2}\left(\sigma^x_j-i\sigma^y_j\right)\prod_{i<j} \sigma^z_i,
    \\
    c_j = \frac{1}{2}\left(\sigma^x_j+i\sigma^y_j\right)\prod_{i<j} \sigma^z_i.
\end{split}
\end{equation}
we can represent the Hamiltonian in terms of Pauli operators
\begin{equation}
    H_{\text{Fermion}} = \frac{h}{2}\sum_{i}\left(\sigma^x_i\sigma^x_{i+1}+\sigma^y_i\sigma^y_{i+1}\right).
\end{equation}

The time evolution operator $U(T) = e^{-iHT}$ can be approximated using Trotter-Suzuki decimation
\begin{equation}
    U(T)\approx \prod_{t=1}^{T/\text{dt}} \Bigg(\prod_i e^{-i \frac{h \text{dt}}{2} \sigma^x_i\sigma^x_{i+1}}e^{-i \frac{h \text{dt}}{2} \sigma^y_i\sigma^y_{i+1}}\Bigg)
\end{equation}
where the error in the approximation is $\mathcal O(\text{dt}^2)$.  
The evolution operator can be written in terms of commonly used quantum gates
\begin{equation}
    U(T)\approx \prod_{t=1}^{T/\text{dt}} \Bigg(\prod_i R^{xx}_{i,i+1}(h \text{dt}) R^{yy}_{i,i+1}(h \text{dt})\Bigg).
\label{UR}
\end{equation}
The order of the products in Eq.~\ref{UR} can be rearranged without incurring error greater than $\mathcal O(\text{dt}^2)$.  Circuits typically perform better if the gates are rearranged as 
\begin{equation}
\begin{split}
    U(T)\approx \prod_{t=1}^{T/\text{dt}} \Bigg( &\prod_{i~\text{odd}} R^{xx}_{i,i+1}(h \text{dt}) R^{yy}_{i,i+1}(h \text{dt}) 
    \\
    \times &\prod_{i~\text{even}} R^{xx}_{i,i+1}(h \text{dt}) R^{yy}_{i,i+1}(h \text{dt}) \Bigg).
\end{split}
\end{equation}

Figure~\ref{fblock}a shows the corresponding quantum circuit.   The circuit can be constructed out of two basic blocks shaded blue and yellow.  Naively, each $R^{xx}$ and $R^{yy}$ gate can be constructed using 2 CNOT gates and 4 single qubit rotations.   Thus, each block contains 12 CNOT gates and the naive complexity of each block is $3\sqrt{3}\pi$.  However, from Fig.~\ref{fblock}b,c we see that the complexity of the entire block can be reduced to less than the complexity of a single CNOT gate.  The complexity of each block is reduced to only $4\%$ of the naive complexity.  This reduction carries over to the the full circuit.  Thus, by building the circuit out of the reduced blocks, we are able to time evolve fermionic chains for many more Trotter steps than would otherwise be possible.  Furthermore, in Fig~\ref{fblock}d,e we see that the error from using the reduced block goes to zero.

\section{Conclusion}
    In conclusion, we have developed a modern python implementation of a method specified in \cite{geo_quant} for determining geodesic control functions of quantum algorithms, called QGeo. Numerical methods must be used because analytic methods that can analyze geodesic complexities for circuits with more than two qubits are not available. Never the less, analytic methods are useful for obtaining complexities of one and two qubit gates which can give context to the full numerical solutions. This implementation has been used to determine whether more efficient versions of quantum algorithms are available. In the case of the quantum fourier transform, this python package has shown speed-ups are available which can reduce complexities by about an order of magnitude as compared with the naive implementation. For quantum simulation of a fermionic chain, QGeo has been used to determine that more efficient implementations of Trotter blocks exist which are $96\%$ more efficient than existing implementations. Therefore, while it is prohibitive to assess circuits with more than six qubits using QGeo, it is still possible to analyze and optimize circuits which have many more qubits, provided that they have a block type structure. In the case of the randomly generated circuit, our procedure was able find an implementation of the circuit which lowered the required amount of effort by two orders of magnitude. 

    How one can implement an actual geodesic algorithm on an existing quantum processor is still not known. However, using the control functions obtained by QGeo and mapping them to existing quantum processors may be possible using more detailed knowledge of the actual processor architecture. This would however require modifications to the super-operator, $\mathcal{G}$, to reflect characteristics of a given processor. These characteristics may include processor topology, error rates, availability of single and two qubit pauli strings, and mappings from RF pulses to pauli strings. 

    Extensions of QGeo which have better scaling properties to more qubits may be possible. This is because the space of tangent vectors with weight less than three only scales quadratically whereas the full tangent space scales exponentially.  However, a realization of better scaling is beyond the scope of this report and we reserve it for future work.

\section*{Acknowledgements} \label{sec:acknowledgements}
    This work was funded by the NISE program at NIWC PAC.  This work has also been supported by the Office of Naval Research (ONR) through the U.S. Naval Research Laboratory (NRL).  We acknowledge QC resources from IBM through a collaboration with the Air Force Research Laboratory (AFRL).

\appendix*
\section{Appendix} \label{sec:appendix}

    Generally solutions the Sch\"odinger equation can be found from:

    \begin{align}
        \label{eq:Schrodinger Sol w/o commuting H}
        U\left(t\right)=1+\sum_{n=1}^{\infty}\left(\frac{-i}{\hbar}\right)^{n}\\\nonumber\intop_{0}^{t}dt_{1}\intop_{0}^{t_{1}}dt_{2}\dots\intop_{0}^{t_{n-1}}dt_{n}H\left(t_{1}\right)H\left(t_{2}\right)\dots H\left(t_{n}\right)
    \end{align}

    This can also be expressed as a time ordered exponential which is a solution because:
    \begin{widetext}
    \begin{equation}\label{solution}
        \begin{split}
            i\frac{d \hat{U}}{dt}&=i \frac{d}{dt}\left(\mathcal{T} \mathrm{e}^{-i \int_{0}^{t}dt' \hat{H}(t')}\right)\\
            &=i \frac{d}{dt}\mathcal{T}\left(1+\sum_{n=1}^{n=\infty}\frac{1}{n!}(-i)^n\int_{0}^{t}dt_1...\int_{0}^{t}dt_n \hat{H}(t_1)... \hat{H}(t_n)\right)\\
            &=i \mathcal{T}\left(\sum_{n=1}^{n=\infty}\frac{1}{n!}(-i)^n\frac{d}{dt}\int_{0}^{t}dt_1...\int_{0}^{t}dt_n \hat{H}(t_1)... \hat{H}(t_n)\right)\\
            &=i \mathcal{T}\left(\sum_{n=1}^{n=\infty}\sum_{i=1}^{i=n}\frac{1}{n!}(-i)^n\int_{0}^{t}dt_1...\int_{0}^{t}dt_{i-1}\int_{0}^{t}dt_{i+1}...\int_{0}^{t}dt_n \hat{H}(t_1)...\hat{H}(t_{i-1})\hat{H}(t)\hat{H}(t_{i+1})... \hat{H}(t_n)\right)\\
            &=\hat{H}(t)+\hat{H}(t)\sum_{n=2}^{n=\infty}\frac{1}{(n-1)!}(-i)^{n-1}\int_{0}^{t}dt_1...\int_{0}^{t}dt_{n-1} \mathcal{T}\left(\hat{H}(t_1)... \hat{H}(t_{n-1})\right)\\
            &=\hat{H}(t)\left(1+\sum_{n=1}^{n=\infty}\frac{1}{(n)!}(-i)^n\int_{0}^{t}dt_1...\int_{0}^{t}dt_n \mathcal{T}\left(\hat{H}(t_1)... \hat{H}(t_n)\right)\right)\\
            &=\hat{H}(t)\hat{U}.\\
        \end{split}
    \end{equation}
    \end{widetext}

    Also, solutions to Schr{\"o}dinger's equation can be simplified through several different assumptions such as: 1. H is independent of time, 2.  H is time-dependent but the H's at different times commute, and the general case, case 3. The H's at different times do not commute.
     Case 1 takes the form:

    \begin{align}
        \label{eq:Schrodinger Sol w const H}
        U\left(t\right)=\exp\left(-\frac{i}{\hbar}\left(t-t_{0}\right)H\right)U_{0}
    \end{align}

    Case 2 takes the form:

    \begin{align}
        \label{eq:Schrodinger Sol w commuting H}
        U\left(t\right)=\exp\left(-\frac{i}{\hbar}\intop_{t_{0}}^{t} H\left(t^{\prime}\right) dt^{\prime}\right)U_{0}
    \end{align}

    Case 3 takes the form:

    \begin{align}
        \label{eq:Schrodinger Sol w/o commuting H}
        U\left(t\right)=1+\sum_{n=1}^{\infty}\left(\frac{-i}{\hbar}\right)^{n}\\\nonumber\intop_{0}^{t}dt_{1}\intop_{0}^{t_{1}}dt_{2}\dots\intop_{0}^{t_{n-1}}dt_{n}H\left(t_{1}\right)H\left(t_{2}\right)\dots H\left(t_{n}\right)
    \end{align}

    We usually will have each of the cases we have beginning at time $t=0$. 

    Finally, we list here the basis elements for Hamiltonians living in $su(2)$, $su(4)$ and $su(8)$. For notational convenience we suppress knronecker products and sigmas. So for example in our notation we have $x=\sigma_x$, and $ x\mathbb{1}z=\sigma_x\otimes\mathbb{1}\otimes \sigma_z$. Given this notation we have the following basis sets:
    \begin{align}
        \label{eq:Metric Basis n=1}
        \scalebox{0.9}{\(
        \left[\begin{array}{ccc}
        x & y & z
        \end{array}\right]
        \)}
    \end{align}

    \begin{align}
        \label{eq:Metric Basis n=2}
        \scalebox{0.9}{\(
        \left[\begin{array}{ccccc}
        \mathbb{1}x & \mathbb{1}y & \mathbb{1}z & x\mathbb{1} & xx,\\
        xy & xz & y\mathbb{1} & yx & yy,\\
        yz & z\mathbb{1} & zx & zy & zz
        \end{array}\right]
        \)}
    \end{align}

    \begin{align}
        \label{eq:Metric Basis n=3}
        \scalebox{0.9}{\(
        \left[\begin{array}{ccccccccc}
        \mathbb{1}\mathbb{1}x & \mathbb{1}\mathbb{1}y & \mathbb{1}\mathbb{1}z & \mathbb{1}x\mathbb{1} & \mathbb{1}xx & \mathbb{1}xy & \mathbb{1}xz & \mathbb{1}y\mathbb{1} & \mathbb{1}yx,\\
        \mathbb{1}yy & \mathbb{1}yz & \mathbb{1}z\mathbb{1} & \mathbb{1}zx & \mathbb{1}zy & \mathbb{1}zz & x\mathbb{1}\mathbb{1} & x\mathbb{1}x & x\mathbb{1}y,\\
        x\mathbb{1}z & xx\mathbb{1} & xxx & xxy & xxz & xy\mathbb{1} & xyz & xyy & xyz,\\
        xz\mathbb{1} & xzx & xzy & xzz & y\mathbb{1}\mathbb{1} & y\mathbb{1}x & y\mathbb{1}y & y\mathbb{1}z & yx\mathbb{1},\\
        yxx & yxy & yxz & yy\mathbb{1} & yyx & yyy & yyz & yz\mathbb{1} & yzx,\\
        yzy & yzz & z\mathbb{1}\mathbb{1} & z\mathbb{1}x & z\mathbb{1}y & z\mathbb{1}z & zx\mathbb{1} & zxx & zxy,\\
        zxz & zy\mathbb{1} & zyx & zyy & zyz & zz\mathbb{1} & zzx & zzy & zzz
        \end{array}\right]
        \)}
    \end{align}

\bibliographystyle{apsrev4-2}
\bibliography{ref}

\afterpage{%
\begin{widetext}
\begin{figure}[b]
    \centering
    \includegraphics[width=\textwidth]{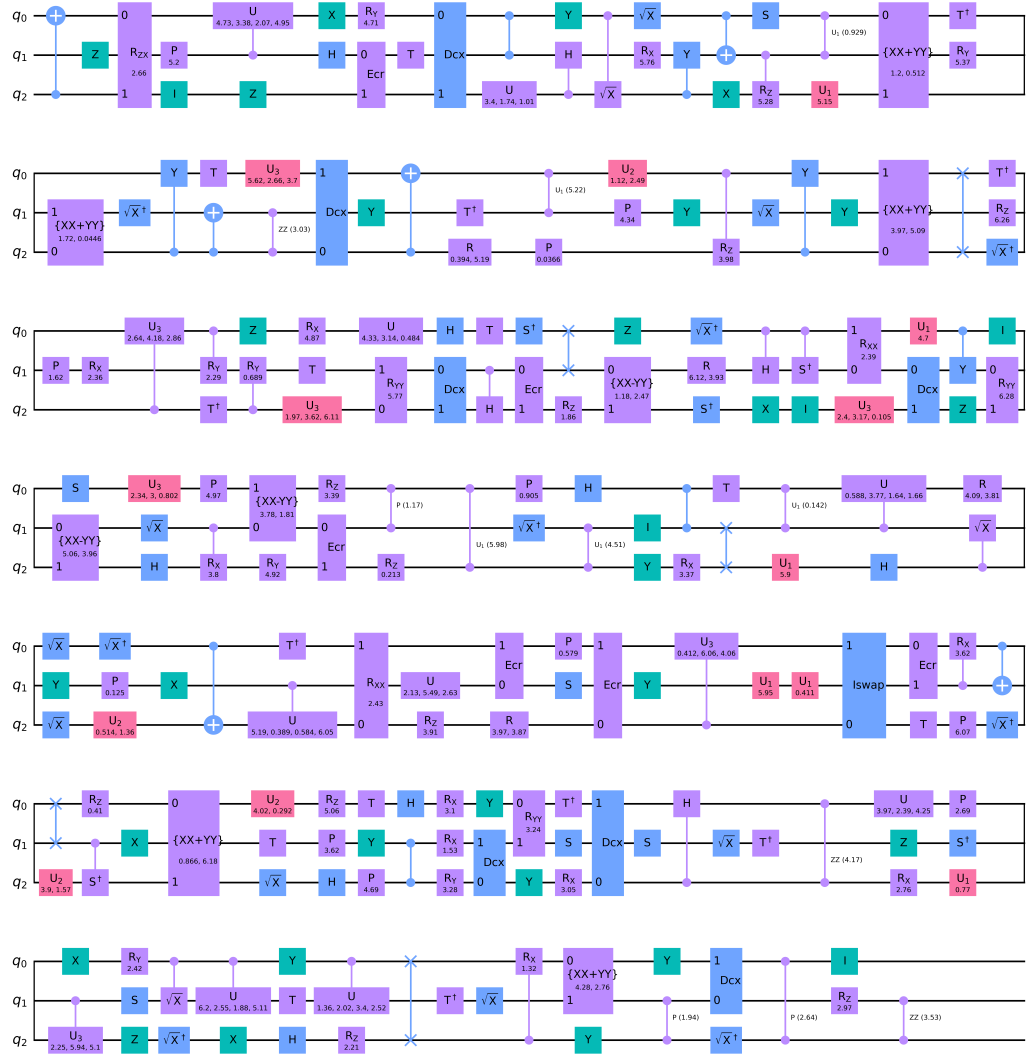}
    \caption{A randomly generated circuit.}
    \label{fig:random_circuit}
\end{figure}

\begin{figure}[b]
    \centering
    \includegraphics[width=\textwidth]{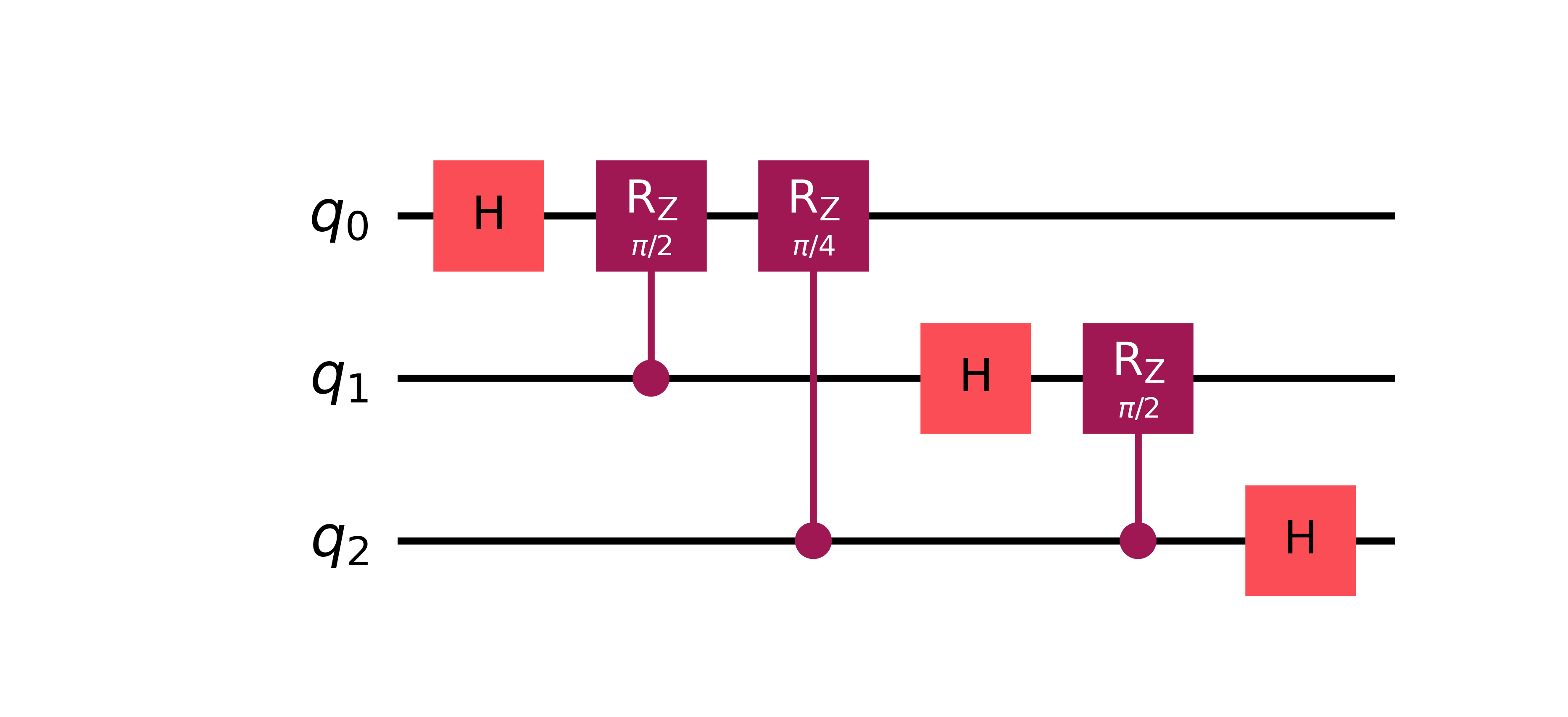}
    \caption{The three qubit quantum fourier transform.}
    \label{fig:3qft}
\end{figure}

\begin{figure}[b]
    \centering
    \includegraphics[width=\textwidth]{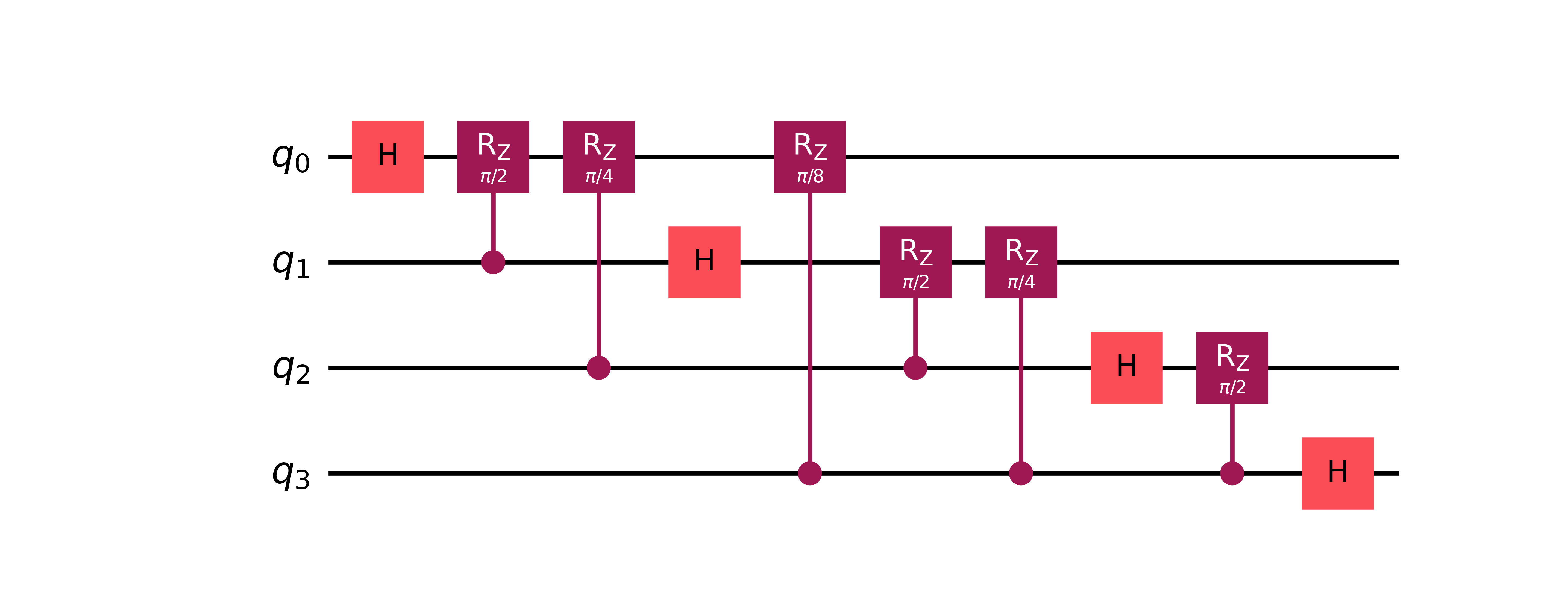}
    \caption{The four qubit quantum fourier transform.}
    \label{fig:4qft}
\end{figure}

\begin{figure}[b]
    \centering
    \includegraphics[width=\textwidth]{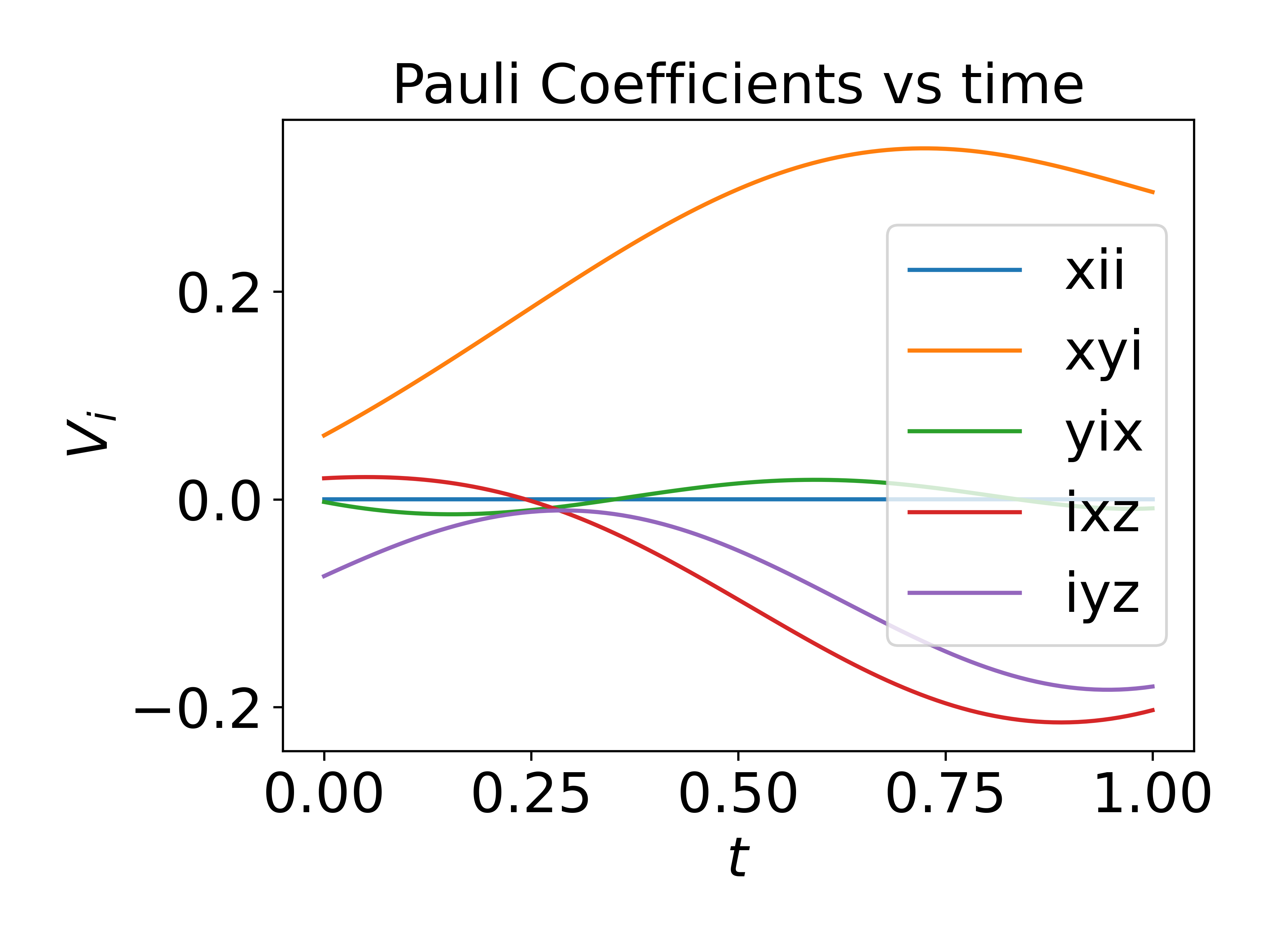}
    \caption{A selection of control functions from \ref{fig:QFT_coeffs_v_t}.}
    \label{fig:labled_control}
\end{figure}

\begin{figure}[b]
    \centering
    \includegraphics[width=\textwidth]{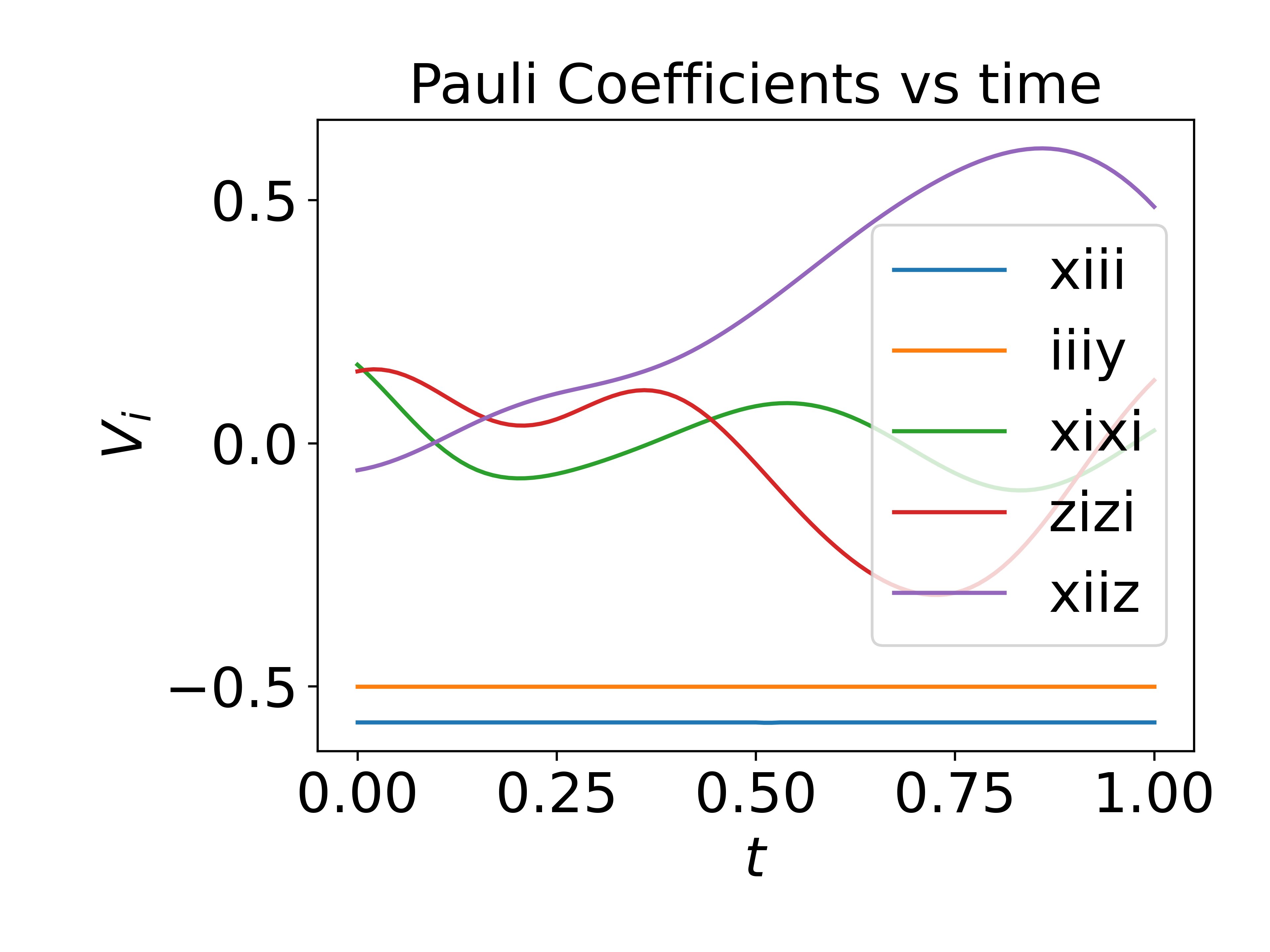}
    \caption{A selection of control functions from \ref{fig:QFT_coeffs_4_v_t}.}
    \label{fig:labled_control_4}
\end{figure}
\end{widetext}}

\end{document}